\documentclass[aps,preprint,nofootinbib,floatfix]{revtex4-1}
\usepackage{graphicx,hyperref}
\usepackage{amsmath,amssymb}
\usepackage{url}
\usepackage{color,epstopdf}
\newcommand{\lsim}{\mathrel{\mathop{\kern 0pt \rlap
  {\raise.2ex\hbox{$<$}}}
  \lower.9ex\hbox{\kern-.190em $\sim$}}}
\newcommand{\gsim}{\mathrel{\mathop{\kern 0pt \rlap
  {\raise.2ex\hbox{$>$}}}
  \lower.9ex\hbox{\kern-.190em $\sim$}}}

\begin{document}
\title{Diphoton rate of the standard-model-Like Higgs boson \\
in the extra $U(1)$ extended MSSM\\}
\author{Kingman Cheung$^{1,2}$, Chih-Ting Lu$^{1}$, and Tzu-Chiang Yuan$^3$}

\affiliation{
$^1$Department of Physics, National Tsing Hua University, 
Hsinchu 300, Taiwan\\
$^2$Division of Quantum Phases \& Devices, School of Physics, 
Konkuk University, Seoul 143-701, Republic of Korea \\
$^3$Institute of Physics, Academia Sinica, Nankang, Taipei 11529, Taiwan
}

\date{\today} 

\begin{abstract}
  Motivated by the excess in the diphoton production rate of the Higgs
  boson at the Large Hadron Collider (LHC), we investigate the
  possibility that one of the CP-even Higgs bosons of the extra
  $U(1)$ extended minimal supersymmetric standard model can
  give a consistent result.  We scan the parameter space for a
  standard-model-like Higgs boson such that the mass is in the range
  of $124-127$ GeV and the production rate $\sigma \cdot B$ of the $W
  W^*$, $ZZ^*$ modes is consistent with the standard model (SM)
  values while that of
  $\gamma\gamma$ is enhanced relative to the SM value.  We find that
  the SM-like Higgs boson is mostly the lightest CP-even Higgs boson
  and it has a strong mixing with the second lightest one, which is
  largely singletlike. 
  The implications on $Z\gamma$ production rate and properties of 
  the other Higgs bosons are also studied.
\end{abstract}

\maketitle

\section{Introduction}

A boson of mass 125 GeV, almost consistent with the standard model
(SM) Higgs boson, was recently discovered at the LHC experiments
\cite{cms,atlas}.  The production rates of $pp \to h \to WW^*, ZZ^*$
are consistent with the SM values while that of $pp \to h \to
\gamma\gamma$ is somewhat higher than the SM expectation. On the other hand,
fermionic modes $b\bar b$ and $\tau\tau$ seem to be  suppressed
 in the 
present data set, however, the uncertainties are still too large to say
anything concrete.  Nevertheless, the diphoton rate has seemed to stay
above the SM prediction since 2011.  If it is confirmed after collecting 
more data at the end of 2012, this would become a strong constraint on 
supersymmetry models and on other extended Higgs models. 

In the minimal supersymmetric standard model (MSSM), the mass of 
the lightest CP-even Higgs boson can be raised to 125 GeV by a large 
radiative correction with a relatively large soft parameter $A_t$ 
\cite{mssm,carena}.
However, the more difficult requirement is to achieve an enhanced
diphoton production rate $gg \to h \to \gamma\gamma$ relative to the
SM prediction.  One possibility is to have a light scalar tau (or stau),
as light as 100 GeV, which is made possible by choosing 
the third generation slepton masses
$m_{L_3}, m_{E_3} \sim 200-450$ GeV, 
the parameter $\mu \sim 200-1000$ GeV, and large
$\tan\beta \sim 60$ \cite{carena}. Such a light stau will soon be 
confirmed or ruled out at the LHC. Another possibility is to identify the
heavier CP-even Higgs boson as the observed 125 GeV boson 
\footnote{
A similar consideration was also analyzed for the two-Higgs-doublet models in
Ref.\cite{Drozd:2012vf}.
}
and the enhancement of the diphoton rate is then made possible by a reduction
in $b\bar b$ width  \cite{mssm-d}. In this case, all the other Higgs bosons 
are around or below 125 GeV, which will soon be uncovered at the LHC.  
Yet, one can also fine-tune the mixing angle ($\alpha$)
between the two CP-even Higgs bosons such that the observed one is mostly
$H_u$, the Higgs doublet that couples to the right-handed up-type quarks. 
In this case, one can achieve enhancement to the diphoton rate
and suppression to the $\tau\tau$ mode \cite{yee}. Among all these
possible scenarios certain levels of fine-tuning are necessary to 
achieve a 125 Higgs boson and an enhanced diphoton rate. 

It is quite well known that the next-to-minimal supersymmetric standard 
model (NMSSM) gives additional tree-level contributions to the Higgs boson
mass arising from the terms $\lambda S H_u H_d$ and $\kappa S^3/3$ in the
superpotential and the corresponding soft terms. The 125 GeV CP-even Higgs
boson can be obtained as either the lightest or the second lightest one, 
without giving stress on the stop sector \cite{ellw,nmssm,kai}.  
The diphoton production
rate can be enhanced through the singlet-doublet mixing. The second lightest
CP-even Higgs boson is the SM-like one while the lightest CP-even is more
singletlike. The mixing between these two states substantially reduces
the $b\bar b$ width of the SM-like Higgs boson, which then causes an
increase of the branching ratio into $\gamma\gamma$ \cite{ellw,nmssm}.

In this work, we consider the $U(1)'$-extended minimal supersymmetric
standard model (UMSSM), which involves an
extra $U(1)$ symmetry and a Higgs singlet superfield $S$.
It is well known that by adding the singlet Higgs field one can easily raise
the Higgs boson mass.
The scalar component of the Higgs singlet
superfield develops a vacuum expectation value (VEV), which 
breaks the $U(1)'$ symmetry and gives a mass to the $U(1)'$ gauge boson,
denoted by $Z'$.  At the same time, the VEV together with the Yukawa
coupling can form an effective $\mu_{\rm eff}$ parameter from the term
$\lambda \langle S \rangle H_u H_d = \mu_{\rm eff} H_u H_d$ in the
superpotential, thus solving the $\mu$ problem of MSSM. Also, because
of the presence of the $U(1)'$ symmetry, terms like $S, S^2$, or $S^3$
are disallowed in the superpotential.

The existence of extra neutral gauge bosons had been predicted in many 
extensions
of the SM \cite{paul}.  String-inspired models and
grand-unified theory (GUT) models usually contain a number of extra $U(1)$
symmetries, beyond the hypercharge $U(1)_Y$ of the SM.  The exceptional group
$E_6$ is one of the famous examples of this type.
Phenomenologically, the most interesting option
is the breaking of these $U(1)$'s at around TeV scales, giving rise to
an extra neutral gauge boson observable 
at the Tevatron and the LHC.
Previously, in the works of Refs.\cite{ours,kang}, a scenario of $U(1)'$ symmetry breaking at
around TeV scale by the VEV of a Higgs singlet superfield in the
context of weak-scale supersymmetry was considered.  The $Z'$ boson obtains a mass from
the breaking of this $U(1)'$ symmetry that is proportional to the VEVs.
Such a $Z'$ can decay into the SUSY particles such as neutralinos,
charginos, and sleptons, in addition to the SM particles.  Thus, the
current mass limits are reduced by a substantial amount and so is the
sensitivity reach at the LHC \cite{ours,kang}. 
We have also considered the SM-like boson and its decay branching ratios
into $WW^*,\; ZZ^*$, and $\tilde{\chi}_1^0 \tilde{\chi}_1^0$  with the
Higgs boson mass in the ranges of $120-130$ and $130-141$ GeV \cite{mimic}.
In the first mass range $120-130$ GeV, we selected the parameter space
such that the SM-like Higgs boson behaves like the SM Higgs boson while in the
second mass range $130-141$ GeV, we selected the parameter space point to
make sure that the Higgs boson is hiding from the existing data.

The goal in this work is to refine the previous analyses \cite{mimic} to scan for the
parameter space such that
\begin{enumerate}
\item the SM-like Higgs boson falls in the mass range $124-127$ GeV;
\item the production rates for $gg \to h \to WW^*, ZZ^*$ are consistent
 with the SM within certainties;
\item the production rate for $gg \to h \to \gamma\gamma$ is enhanced 
relative to the SM prediction, namely,
 \[
   R_{\gamma\gamma} \equiv \frac{\sigma(gg \to h) \times B(h \to \gamma\gamma)}
        {\sigma(gg \to h_{\rm SM}) \times B(h_{\rm SM} \to \gamma\gamma)} 
 > 1 \; ;
\]
\item other existing constraints such as $Z$ invisible width and
chargino mass bound are fulfilled. 
\end{enumerate}
In the chosen parameter space, we calculate the $Z\gamma$ production rate
and study the properties of the other Higgs bosons. 

We organize the paper as follows. In the next section, we briefly describe
the model (UMSSM) and summarize the formulas for the one loop decays of the
CP-even Higgs bosons.
In Sec. III, we search for the parameter space in the model that satisfies
the above requirements, and present the numerical results.
We discuss and conclude in Sec. IV. Detailed expressions for the loop functions 
in the decay formulas are relegated to the Appendix.
Some recent studies on extended MSSM can be found in Ref.~\cite{others}
and on extended electroweak models in Ref.~\cite{others1}.

\section{UMSSM}

For illustrative purposes we use the popular grand unified models 
based on the exceptional group  $E_6$, which is anomaly free.
The two most studied $U(1)$ subgroups in the symmetry breaking chain 
of $E_6$ are
\[
  E_6 \to SO(10) \times U(1)_\psi\,, \qquad 
 SO(10) \to SU(5) \times U(1)_\chi  
\]
In $E_6$ each family of the left-handed fermions 
is embedded into a fundamental $\mathbf{27}$-plet, which decomposes under 
$E_6 \to SO(10) \to SU(5)$ as 
\[
\mathbf{27} \to \mathbf{16} + \mathbf{10} + \mathbf{1} \to
  ( \mathbf{10} + \mathbf{5^*} + \mathbf{1} ) + (\mathbf{5} +
  \mathbf{5^*} ) + \mathbf{1} 
\]
The SM fermions of each family together with an extra state identified 
as the conjugate of a right-handed neutrino
are embedded into the $\mathbf{10}$, $\mathbf {5^*}$, and $\mathbf{1}$
of the $\mathbf{16}$.
All the other states are exotic states required for the $\mathbf{27}$-plet of $E_6$ unification.
In general, the two $U(1)_\psi$ and $U(1)_\chi$ are allowed to mix as
\begin{equation}
  Q'(\theta_{E_6} ) = \cos \theta_{E_6} Q'_\chi + \sin \theta_{E_6} Q'_\psi \;,
\end{equation}
where $0 \le \theta_{E_6} < \pi$ is the mixing angle.  The commonly studied
$Z'_\eta$ model assumes the mixing angle  
$\theta_{E_6} = \pi - \tan^{-1} \sqrt{5/3} \sim 0.71 \pi$ such that
\begin{equation}
\label{eta-model}
 Q'_\eta = \sqrt{ \frac{3}{8} } Q'_\chi - \sqrt{ \frac{5}{8} } Q'_\psi \;.
\end{equation}
Here we follow the common practice by assuming that 
all the exotic particles, other than the 
particle contents of the MSSM, are very heavy and well beyond the reaches
of all current and planned colliders. For an excellent review of $Z'$ models,
see Ref. \cite{paul}.

The effective superpotential $W_{\rm eff}$ involving the matter and Higgs superfields in UMSSM can be written as 
\begin{equation}
\label{sp}
W_{\rm eff} = \epsilon_{ab} \left [ y^u_{ij} Q^a_j H_u^b U^{\rm c}_i 
  - y^d_{ij} Q^a_j H_d^b D^{\rm c}_i 
  -  y^l_{ij} L^a_j H_d^b E^{\rm c}_i 
  +  h_s S  H_u^a H_d^b \right ] \;,
\end{equation}
where $\epsilon_{12}= - \,\epsilon_{21} =1$, $i,j$ are family indices,
and $y^u$ and $y^d$ represent the Yukawa matrices for the 
up-type and down-type quarks respectively.
Here $Q, L, U^{\rm c}, D^{\rm c}, E^{\rm c}, H_u$, and $H_d$ denote the MSSM superfields for the 
quark doublet, lepton doublet,
up-type quark singlet, down-type quark singlet, lepton singlet,
up-type Higgs doublet, and down-type Higgs doublet respectively,
and $S$ is the singlet superfield. 
The $U(1)'$ charges of the fields
$H_u, H_d,$ and $S$ are chosen such that the relation 
$Q'_{H_u} + Q'_{H_d} + Q'_S = 0$ holds. Thus
$S H_u H_d$ is the only term in the superpotential allowed by the $U(1)'$ symmetry 
beyond the MSSM. Once the singlet scalar field $S$ develops a 
VEV, it generates an effective $\mu$ parameter: $\mu_{\rm eff} = 
h_s \langle S \rangle$.  

The singlet superfield will give rise to a singlet scalar boson and a
singlino. The real part of the scalar boson will mix with the real
part of $H_u^0$ and $H_d^0$ to form three  physical CP-even Higgs bosons.  The
imaginary part of the singlet scalar will be eaten and 
become the longitudinal part of the $Z'$
boson according to the Higgs mechanism 
in the process of spontaneous symmetry breaking of $U(1)'$.  
The singlino, together with the $Z'$-ino, will mix with the
neutral gauginos and neutral Higgsinos to form six physical neutralinos.
Studies of various singlet extensions of the MSSM can be found in 
Refs.~\cite{vernon-n,vernon-h,sy}.

The Higgs doublet and singlet fields are 
\begin{equation}
H_d = \left( \begin{array}{c}
               H_d^0 \\
               H_d^- \end{array}  \right ) \;\; , \qquad 
H_u = \left( \begin{array}{c}
               H_u^+ \\
               H_u^0 \end{array}  \right )  \;\; \qquad {\rm and} \qquad 
S  \;.
\end{equation}
The scalar interactions are obtained by calculating the 
$F$- and $D$-terms of the superpotential, and by including the 
soft-SUSY-breaking terms.  They are given in Refs.~\cite{ours,mimic}.

Now we can expand the Higgs fields after taking on VEVs as 
\begin{eqnarray}
H_d^0 &=& \frac{1}{\sqrt{2}} \, \left( v_d + \phi_d + i \chi_d \right)
         \,,\nonumber\\
H_u^0 &=& \frac{1}{\sqrt{2}} \, \left( v_u + \phi_u + i \chi_u \right )
   \,,\nonumber\\
S  &=& \frac{1}{\sqrt{2}} \, \left( v_s + \phi_s + i \chi_s \right )
       \,.\nonumber
\end{eqnarray}
It is well known that the lightest CP-even Higgs boson mass receives
a substantial radiative mass correction in the 
MSSM.  The same is true here for the
UMSSM. Tree-level and radiative corrections to the mass matrix 
${\cal M}^{\rm tree}$ have been given in Ref.~\cite{vernon-h}. 
We have included radiative corrections in our calculation.
The interaction eigenstates $\phi_u, \phi_d, \phi_s$ can be rotated into
mass eigenstates via an orthogonal matrix $O$
\begin{equation}
\left( \begin{array}{c} 
               h_1 \\
               h_2 \\
               h_3  \end{array} \right ) =  O \,
 \left( \begin{array}{c} 
                    \phi_d \\
                    \phi_u \\
                     \phi_s  \end{array} \right )  \qquad \;,
\end{equation}
such that $O{\cal M}^{\rm tree + loop}O^T = {\rm diag}( m^2_{h_1},\;
m^2_{h_2}, \; m^2_{h_3} )$ in ascending order.
There are also one CP-odd Higgs boson and a pair of charged Higgs bosons,
as in the MSSM.
Note that the Higgs  boson masses receive extra
contributions from the $D$-term of the $U(1)'$ symmetry (proportional
to $g_2$) and from the $F$-term for the mixing of the doublets with 
the singlet Higgs field (proportional to $h_s$). 

\subsection{Formulas for one loop decays of the CP-even Higgs bosons}

We will present the relevant formulas for the one loop processes of 
$h_j \to \gamma\gamma, Z\gamma$ and $gg$. 
The $gg$ width is relevant for the gluon-fusion production cross section.
The couplings of the neutral CP-even Higgs bosons  
with the SM gauge bosons and fermions, charged Higgs bosons, sfermions, 
charginos and neutralinos have been given in Refs.~\cite{ours,mimic}.

The $\gamma\gamma$ partial decay width of the CP-even Higgs boson ($h_j , \, j=1,2,3$)
receives contributions from all charged particles running in the loop. 
It is given by
\begin{eqnarray}
\label{hgam}
\Gamma (h_j \rightarrow \gamma\gamma)&=& \frac{\alpha^{2}m_{h_j}^{3}}{256\pi^{3}v^{2}}
\left | F_{\tau} + 3 \left( \frac{2}{3} \right)^{2} F_{t}+ 
       3 \left(-\frac{1}{3}\right)^{2} F_{b} +  F_{W}+ F_{h^{\pm}} \right. \nonumber \\
    && + \, F_{\tilde \tau} + \, 3 \left(\frac{2}{3}\right)^{2} F_{\tilde t} 
 \left.  +3 \left(- \frac{1}{3} \right)^{2} F_{\tilde b} + F_{\tilde \chi^{\pm}} \right |^{2} \;,
\end{eqnarray}
where the factor $3$ in front of $F_t, F_b, F_{\tilde t}$, and 
$F_{\tilde b}$ accounts for the 
color factor, and $v^2 = v_u^2 + v_d^2$.  
The expressions for the loop functions $F$
are given in the Appendix. For the decay $h_j \to gg$ where only colored 
particles are
running in the loop, we have
\begin{equation}
\label{hglue}
\Gamma (h_j \rightarrow gg) = \frac{\alpha_{s}^{2}m_{h_j}^{3}}{128\pi^{3} v^{2}}
  \biggl\vert F_{t}+F_{b}+F_{\tilde t}+F_{\tilde b} \biggr\vert^{2}  \;.
\end{equation}
For the decay $h_j \to Z \gamma$, we have
\begin{eqnarray}
\label{hZgam}
\Gamma (h_j \rightarrow Z\gamma) & =& 
\frac{m_{h_j}^{3}}{32\pi}
\left(1-\frac{m_{Z}^{2}}{m_{h_j}^{2}}\right)^{3}
\frac{\alpha ^{2}g^{2}} {16\pi ^{2}m_{W}^{2}} \nonumber \\
&& \times \biggl\vert G_{\tau}+G_{t}+G_{b}+G_{W}+G_{h^{\pm}}
 +G_{\tilde \tau}+G_{\tilde t}+G_{\tilde b} + G_{\tilde \chi ^{\pm}} \biggr\vert^{2} \; .
\end{eqnarray}
The expressions for the loop functions $G$ are given in the Appendix.

\section{Scanning of Parameter Space}

The UMSSM has the following parameters: $M_{\tilde{Z}'}$, $A_s$, 
the VEV $ \langle S \rangle = v_s/\sqrt{2}$, and the Yukawa coupling $h_s$,
other than those of the MSSM: gaugino masses $ M_{1,2,3}$, 
squark masses $M_{\tilde{q}}$, slepton masses $M_{\tilde{\ell}}$, soft 
parameters $A_{t,b,\tau}$, and $\tan\beta$.  The soft parameter
$M_S$ can be expressed in terms of VEVs and couplings through the
tadpole conditions.
The effective $\mu$ parameter is given as
$\mu_{\rm eff} = h_s \langle S \rangle$.  The other model parameters
are fixed by the quantum numbers $Q'_{\phi}$ of various supermultiplets
$\phi$. 

The mass of the $Z'$ boson is determined by 
$m_{Z'} \approx {g_2} (Q'^2_{H_u} v_u^2 +Q'^2_{H_d} v_d^2 + Q'^2_{S} v_s^2 )^{1/2}$
if the $Z-Z'$ mixing is ignored.
The most stringent limit on the $Z'$ boson comes from the dilepton 
resonance search by ATLAS \cite{atlas}. 
Nevertheless, we can avoid 
these $Z'$ mass limits by assuming that
the leptonic decay mode is suppressed.
The mixing between the SM $Z$ boson and the $Z'$ can be suppressed
by carefully choosing the $\tan\beta \approx (Q'_{H_d}/Q'_{H_u})^{1/2}$ 
\cite{vernon-h}.
In this work we do not impose these constraints in our parameter scan.
However we note that we can always carefully choose the set of quantum 
numbers $Q'$ such that both the $Z'$ mass and mixing constraints can be
evaded.
\footnote{Such a $Z'$ boson is still subjected to the dijet resonance
searches.  The CMS Collaboration has published a search for dijet
resonances \cite{cms-dijet}, one of which is the $Z'$ model with the
SM couplings. The production cross section curve of the $Z'$ barely
touches the upper-limit curve and thus receives no constraint. The $Z'$
boson in our case has a smaller coupling down by $g_2/g_1 \approx
0.62$, and so the production cross section is down by $(0.62)^2 =
0.38$. Similarly, it is true for the dijet resonance search in the
mass range $260-1400$ GeV by the CDF Collaboration \cite{cdf-dijet},
which ruled out a part of this $Z'$ mass range when the $Z'$ has the SM
couplings. Again, in our case when the production cross section is
down by $0.38$, the constraint is moot. 
For an even lower mass range of dijet resonance search the relevant
data came from UA2. However, for the $Z'$ model with a coupling
$g_2/g_1 = 0.62$ it has been shown to be safe with the UA2 data in
\cite{song}.  }

We first fix most of the MSSM parameters (unless stated otherwise):
\begin{eqnarray}
&& M_{1} = M_{2} / 2 = 0.2 \;{\rm TeV}, \;\; M_3 = 2 \; {\rm TeV} \; ; \nonumber \\
&& M_{\tilde{Q}} = 0.7 \;{\rm TeV},\;
M_{\tilde{U}} = 0.7 \;{\rm TeV},\; M_{\tilde{D}} = 1 \;{\rm TeV},\;
M_{\tilde{L}} = M_{\tilde{E}} = 1 \;{\rm TeV}\; ;  \\
&&A_b = A_{t} = A_\tau = 1\;{\rm TeV} \; . \nonumber
\end{eqnarray}
We also fix the UMSSM parameter:
\begin{equation}
A_s = 0.5 \, {\rm TeV} \;\; ,
\end{equation}
while we scan the rest of the parameters in the following ranges
\begin{equation}
    0.2 < h_s < 0.6 ,\;\;  1.1 < \tan \beta < 40\; , \;
\label{scan1}
\end{equation}
and
\begin{equation}
0.2 \,{\rm TeV} <  v_s  < 2 \;{\rm TeV},\;\;
0.2 \, {\rm TeV} < M_{\tilde{Z}'}  <  2 \, {\rm TeV} \; \; .
\label{scan2}
\end{equation}
Note that the $U(1)'$ gaugino mass, $M_{\tilde{Z}'}$, is a 
soft-SUSY-breaking parameter, unlike the $Z'$ boson mass which is fixed
by the $U(1)'$ coupling constant and quantum numbers, and the three VEVs.

\subsection{Constraints}

{\it Charginos mass.--} The chargino sector of the UMSSM is the same as
that of MSSM with the following chargino mass matrix
\begin{equation}
 M_{\tilde{\chi}^\pm} = \left ( \begin{array}{cc}
        M_2 & \sqrt{2} m_W \sin\beta \\
        \sqrt{2} m_W \cos\beta & \mu_{\rm eff} \end{array} \right ) \;.
\end{equation}
Thus, the two charginos masses depend on $M_2$,
$\mu_{\rm eff}=h_s v_s/\sqrt{2}$, and $\tan\beta$. 
The current bound for the lighter chargino mass 
is $m_{\tilde{\chi_1}^\pm}> 94$ GeV as long as
its mass difference with the lightest supersymmetric particle (LSP) 
is larger than 3 GeV \cite{pdg}. 
We impose this chargino mass bound in our scans in the parameter space
defined by (\ref{scan1}) and (\ref{scan2}).

{\it Invisible width of the $Z$ boson.--} The lightest neutralino 
$\tilde{\chi}^0_1$ is the LSP of the model, and thus would be stable
and invisible.  When the $Z$ boson decays into a pair of LSPs, it 
would  give rise to an invisible width of the $Z$ boson, which had been
tightly constrained by experiments. The current bound of the $Z$ invisible
width is $\Gamma_{\rm inv} (Z) < 3$ MeV at about 95\% C.L. \cite{pdg}.
The coupling of the $Z$ boson to the lightest neutralino is given by
\begin{equation}
{\cal L}_{Z\tilde{\chi}_1^0 \tilde{\chi}^0_1} = 
  \frac{g_1}{4} \, \left( | N_{13}|^2 - | N_{14} |^2 \right ) \,
Z_\mu \, \overline{\tilde{\chi}^0_1} \gamma^\mu \gamma_5
 \, \tilde{\chi}^0_1\; ,
\end{equation}
where $N$ is the orthogonal matrix that diagonalized the neutralino mass matrix.
The contribution to the $Z$ boson invisible width is
\begin{equation}
\Gamma(Z \to \tilde{\chi}^0_1 \tilde{\chi}^0_1 ) =
  \frac{g_1^2} { 96 \pi} \left( | N_{13}|^2 - | N_{14} |^2 \right )^2  m_Z
 \left( 1 - \frac{4 m_{\tilde{\chi}^0_1}^2 } { m_Z^2 } \right )^{3/2} \; .
\end{equation}
Note that the $Z$ boson would not couple to the singlino 
component, and we have assumed negligible mixing between $Z$ and $Z'$ bosons;
therefore the $Z$ boson would not couple to the $Z'$-ino component either.
Here we impose the experimental constraint on the invisible $Z$ width.
The constraint of fulfilling the relic density by the LSP will be ignored
in this work.

{\it Mass of the Higgs boson and production rate of various decay modes.--}
The boson masses reported by CMS and ATLAS are $125.3 \pm 0.6$ \cite{cms}
and $126.0 \pm 0.6$ GeV \cite{atlas}, respectively. 
The current data indicated that the observed boson is similar to the
SM Higgs boson.
For our purpose we define the SM-like Higgs boson $h_{\rm SM-like}$ 
in our scenario when the square of its 
singlet component is smaller than $1/3$, i.e., $O_{k3}^2 < \frac{1}{3}$, where
$h_k = O_{k1} \phi_d + O_{k2} \phi_u + O_{k3} \phi_s$. 
For all the allowed points we have $k=1$ for the SM-like Higgs boson.
We choose the allowable mass range
for the SM-like Higgs boson in our analysis as
\begin{equation}
  124 \; {\rm GeV} < m_{h_{\rm SM-like}} < 127 \; {\rm GeV} \;.
\end{equation}

The production rate of various channels of the Higgs boson relative to the 
SM prediction is defined as
\begin{equation}
\label{R}
R_{ab} \equiv \frac{ \sigma(pp \to h+X ) \times B(h \to ab)}
             { \sigma(pp \to h_{\rm SM} +X ) \times B(h_{\rm SM} \to ab)}
\end{equation}
where $ab = \gamma\gamma, W^+ W^-, ZZ, b\bar b, \tau^+ \tau^-$. At the LHC,
the production of $h_{\rm SM}$ or the CP-even Higgs bosons in the 
UMSSM is dominated by
 gluon fusion. We shall focus on gluon fusion in Eq.~(\ref{R}).
The production rates of $WW^*$ and $ZZ^*$ reported by CMS and ATLAS are 
close to the SM predictions:
\begin{eqnarray}
 0.2 < R_{WW^*} <1.1 \;, \;\;   0.4<  R_{ZZ^*}  <1.2 & \qquad {\rm CMS} \nonumber \\
 0.8 < R_{WW^*} <1.7 \;, \;\;   0.6<  R_{ZZ^*}  <1.8 & \qquad {\rm ATLAS} \nonumber
\end{eqnarray}
On the other hand, the diphoton production rates reported by CMS \cite{cms}
and ATLAS \cite{atlas} are 
\begin{eqnarray}
1.1 < R_{\gamma\gamma} <2.0 \; , \nonumber \\
1.3 < R_{\gamma\gamma} < 2.2  \;, \nonumber
\end{eqnarray}
respectively. We require in our scan
\begin{eqnarray}
&&   0.5 < R_{WW^*}, R_{ZZ^*} < 1.5 \; , \nonumber \\
&&   1.0 < R_{\gamma\gamma}  \; .
\end{eqnarray}

Current limits on the pseudoscalar Higgs boson ($A$) come from the LEP searches
in the associated production with a scalar Higgs boson ($H$)
of $e^+ e^- \to Z^* \to AH$.
In those MSSM-extended models, such as
NMSSM, where multiple scalar and pseudoscalar Higgs bosons exist, 
the constraint could be severe.
However, there is only one pseudoscalar Higgs boson in the UMSSM
and in our choice of parameters it is often heavier than a few hundred GeV.
Thus, it is not constrained by the current limits.  Similarly, the charged
Higgs boson is also heavy and not constrained by current searches.

\subsection{Numerical results}

\begin{figure}[th!]
\centering
\includegraphics[angle=270,width=3.2in]{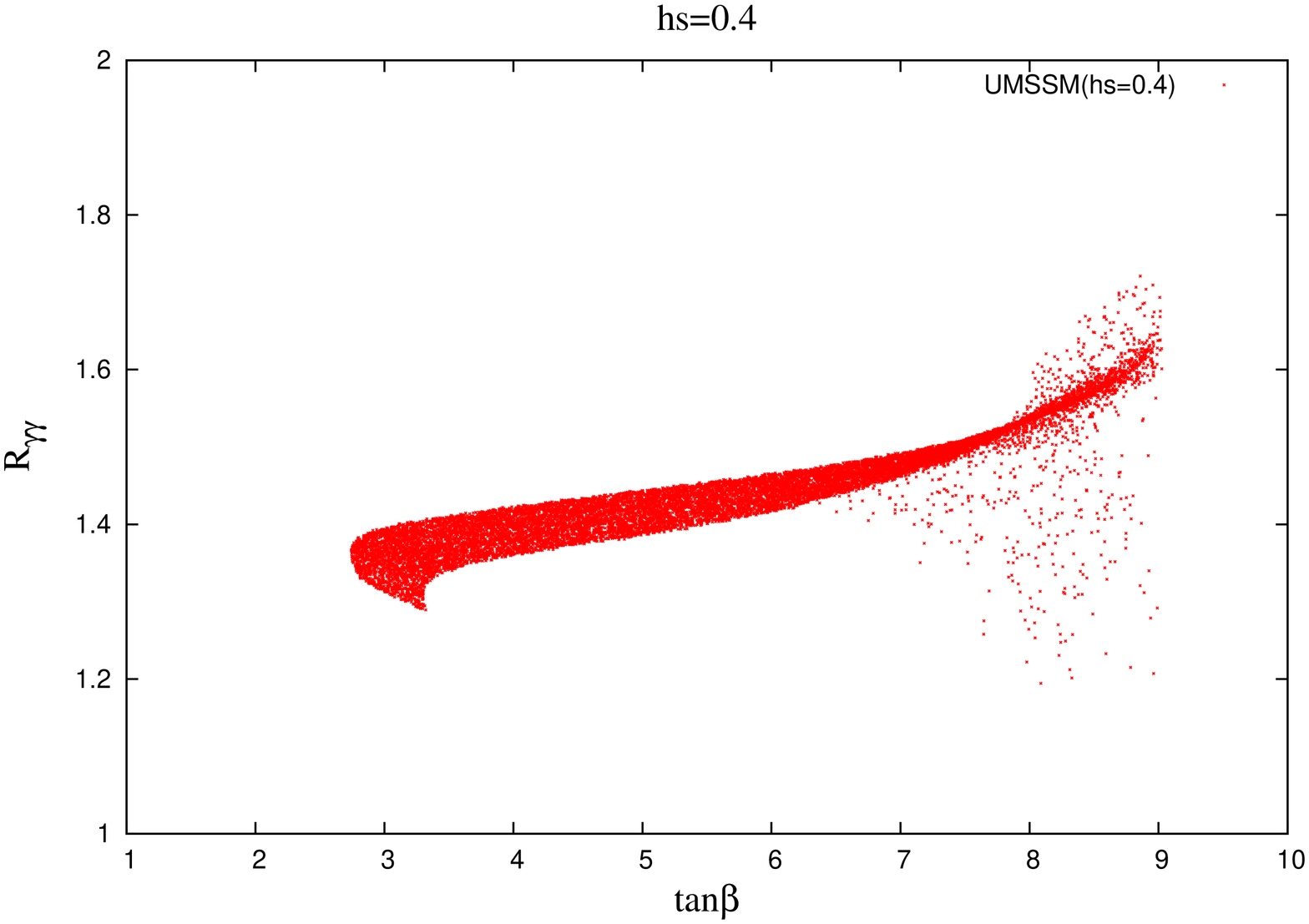}
\includegraphics[angle=270,width=3.2in]{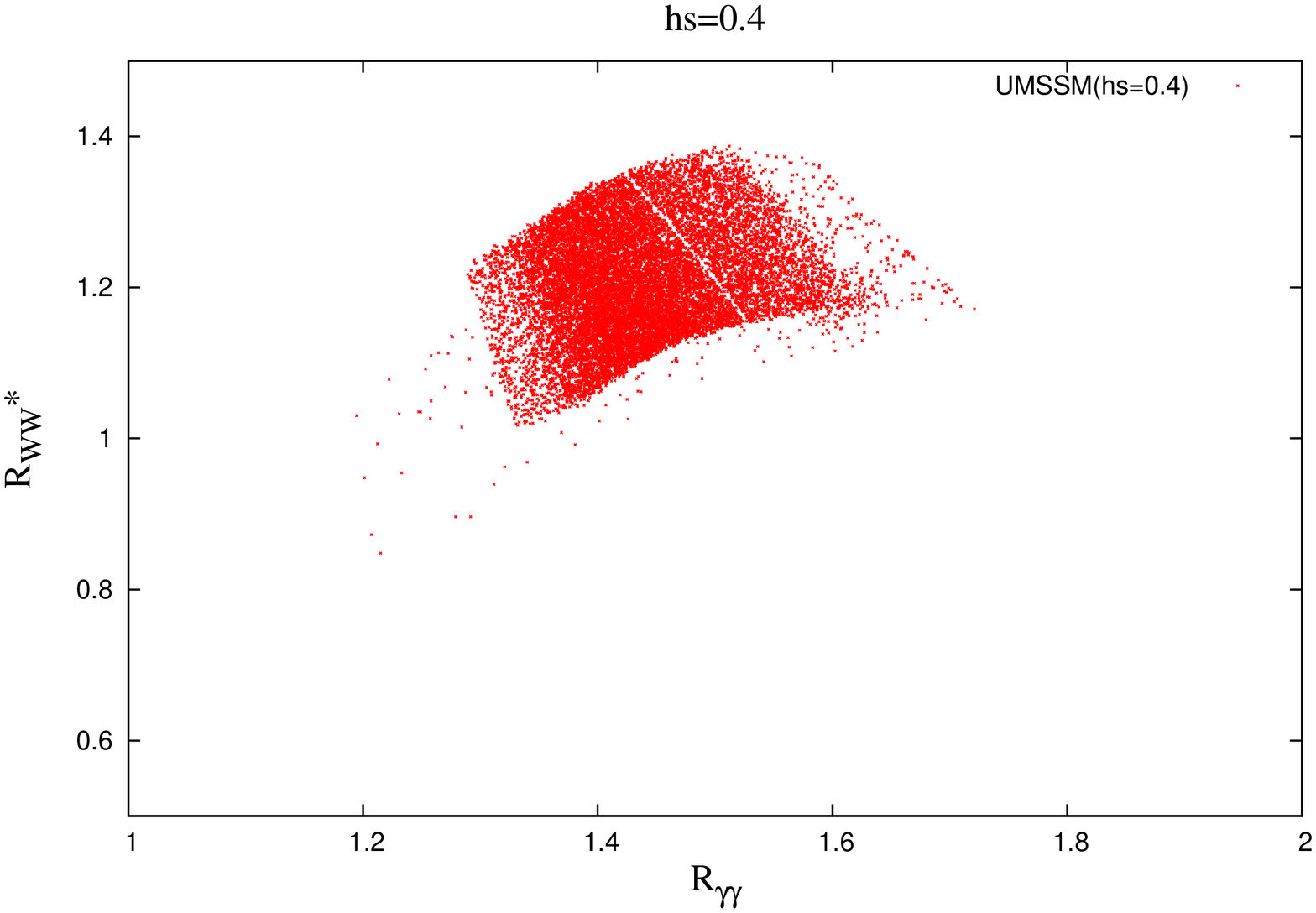}
\includegraphics[angle=270,width=3.2in]{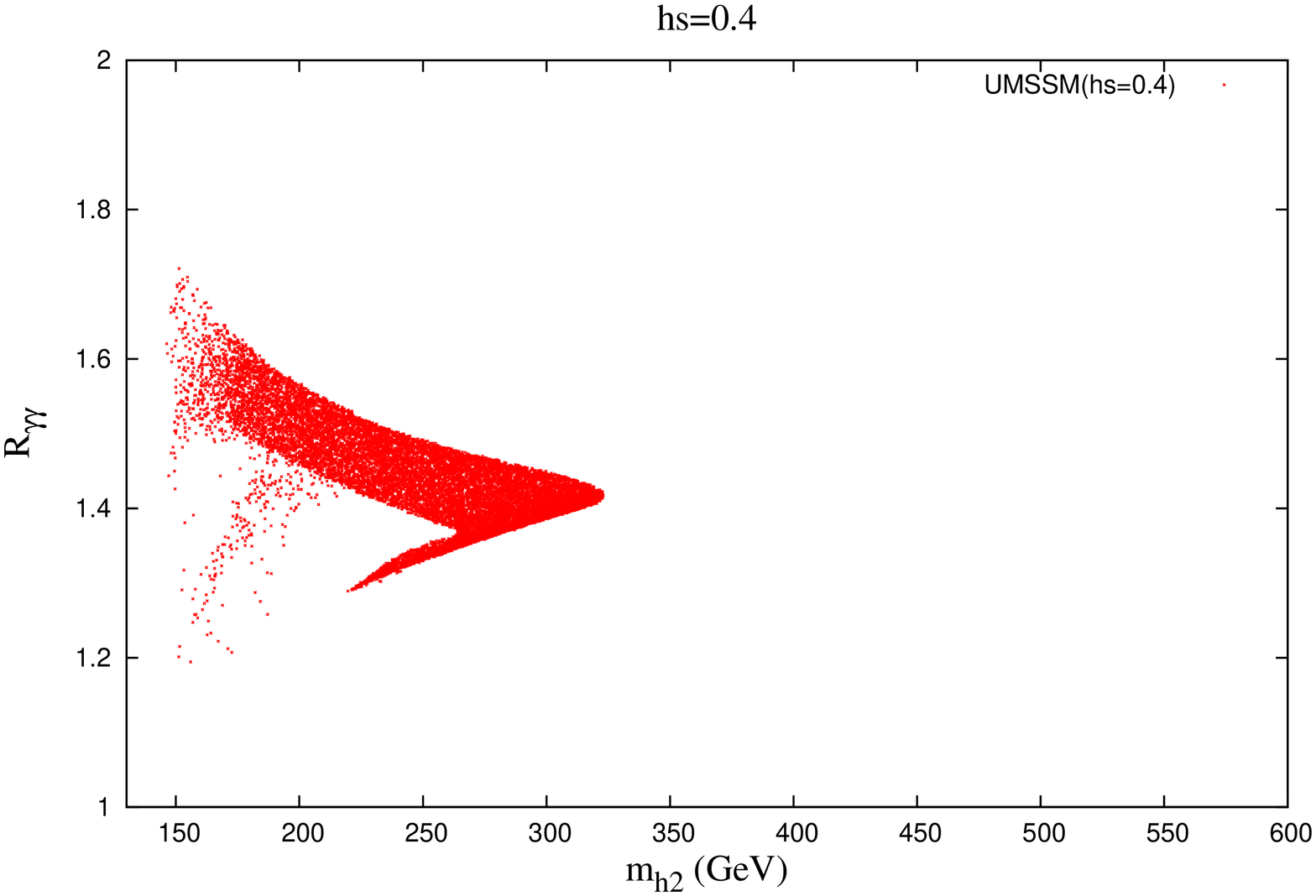}
\includegraphics[angle=270,width=3.2in]{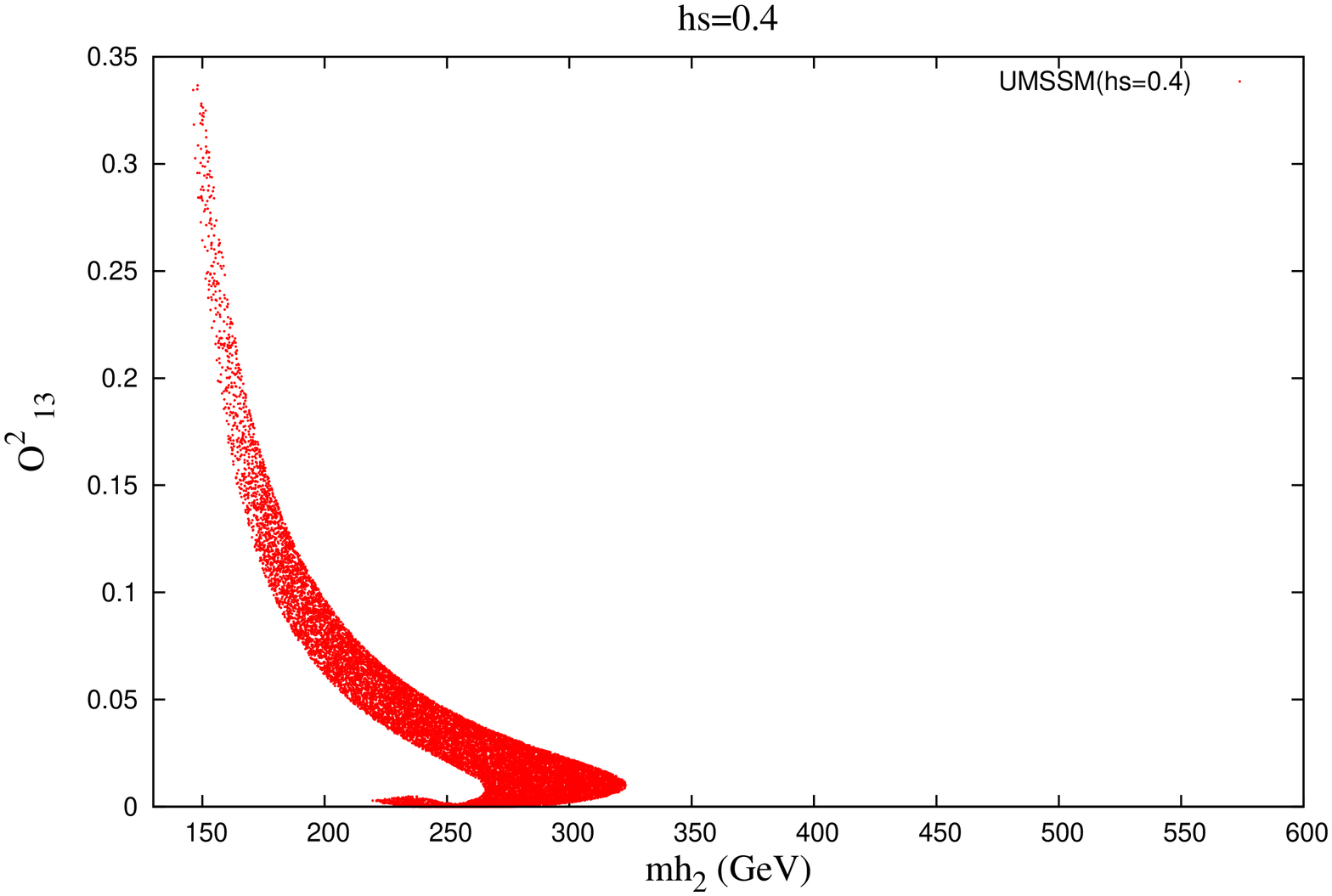}
\caption{\small \label{one}
Case for $h_s = 0.4$. 
Parameter space points satisfy $124\;{\rm GeV} < m_{h_{\rm SM-like}} <
127 \;{\rm GeV}$, the chargino mass, and the invisible $Z$ width constraints.
Also, the relative production rates satisfy
$0.5 < R_{WW^*,ZZ^*} < 1.5$ and $ 1 < R_{\gamma\gamma}$.
 }
\end{figure}

\begin{figure}[th!]
\centering
\includegraphics[angle=270,width=3.2in]{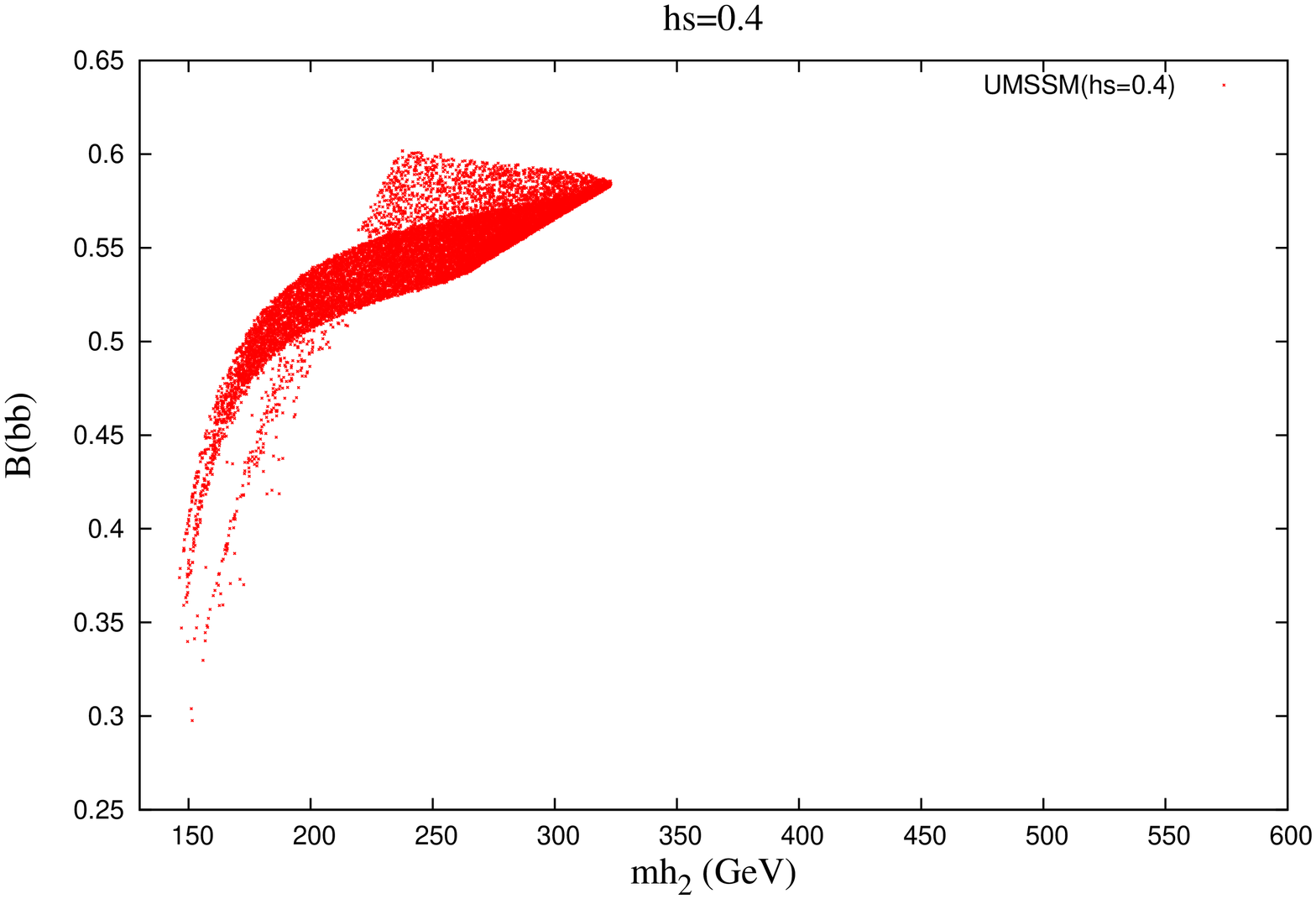}
\includegraphics[angle=270,width=3.2in]{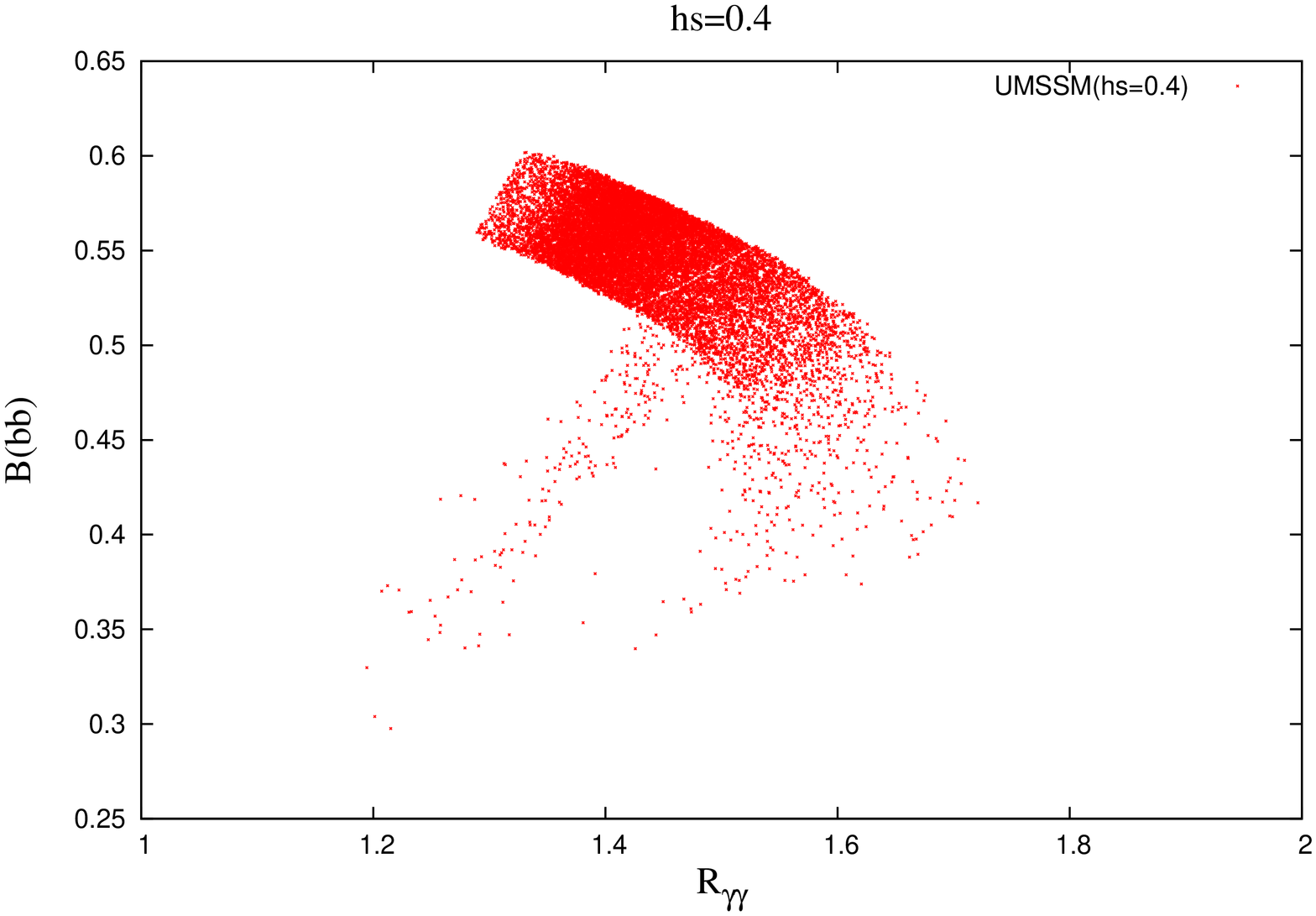}
\caption{\small \label{new2}
Case for $h_s = 0.4$. 
Same as Fig.~\ref{one}, but showing $B(h_1 \to b \bar b)$ versus
(a) $m_{h_2}$ and (b) $R_{\gamma\gamma}$.}
\end{figure}

\begin{figure}[th!]
\centering
\includegraphics[angle=270,width=3.2in]{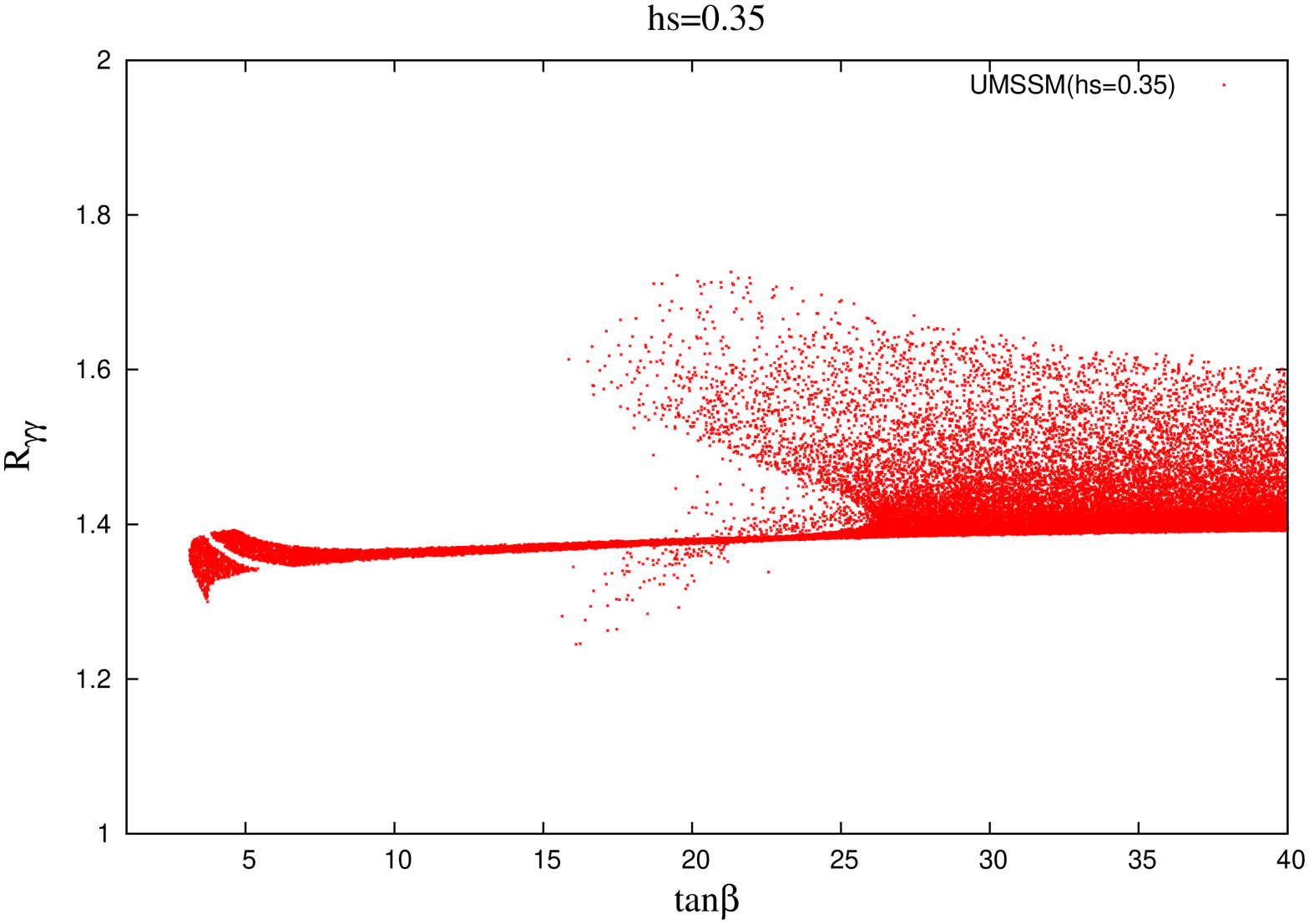}
\includegraphics[angle=270,width=3.2in]{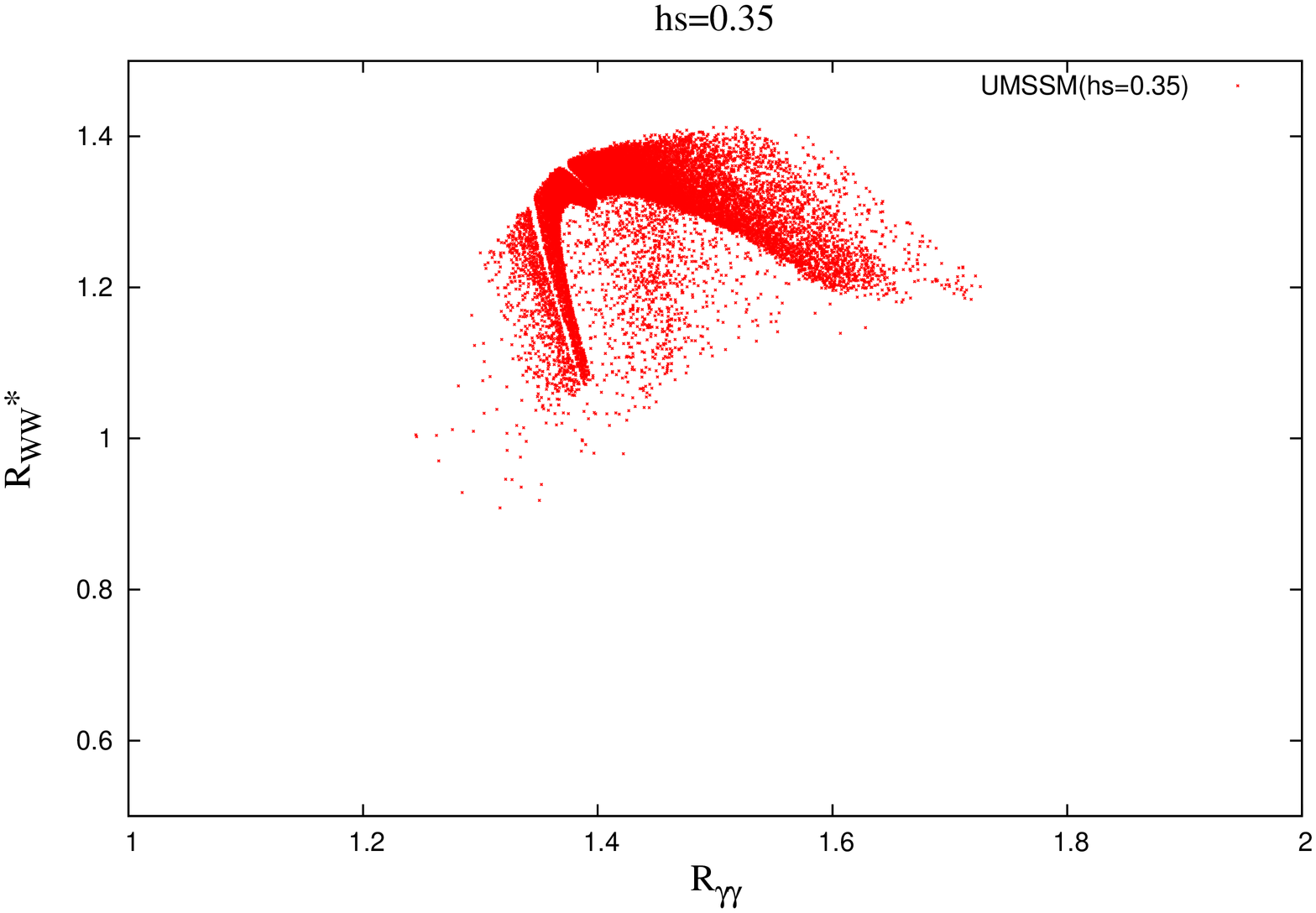}
\includegraphics[angle=270,width=3.2in]{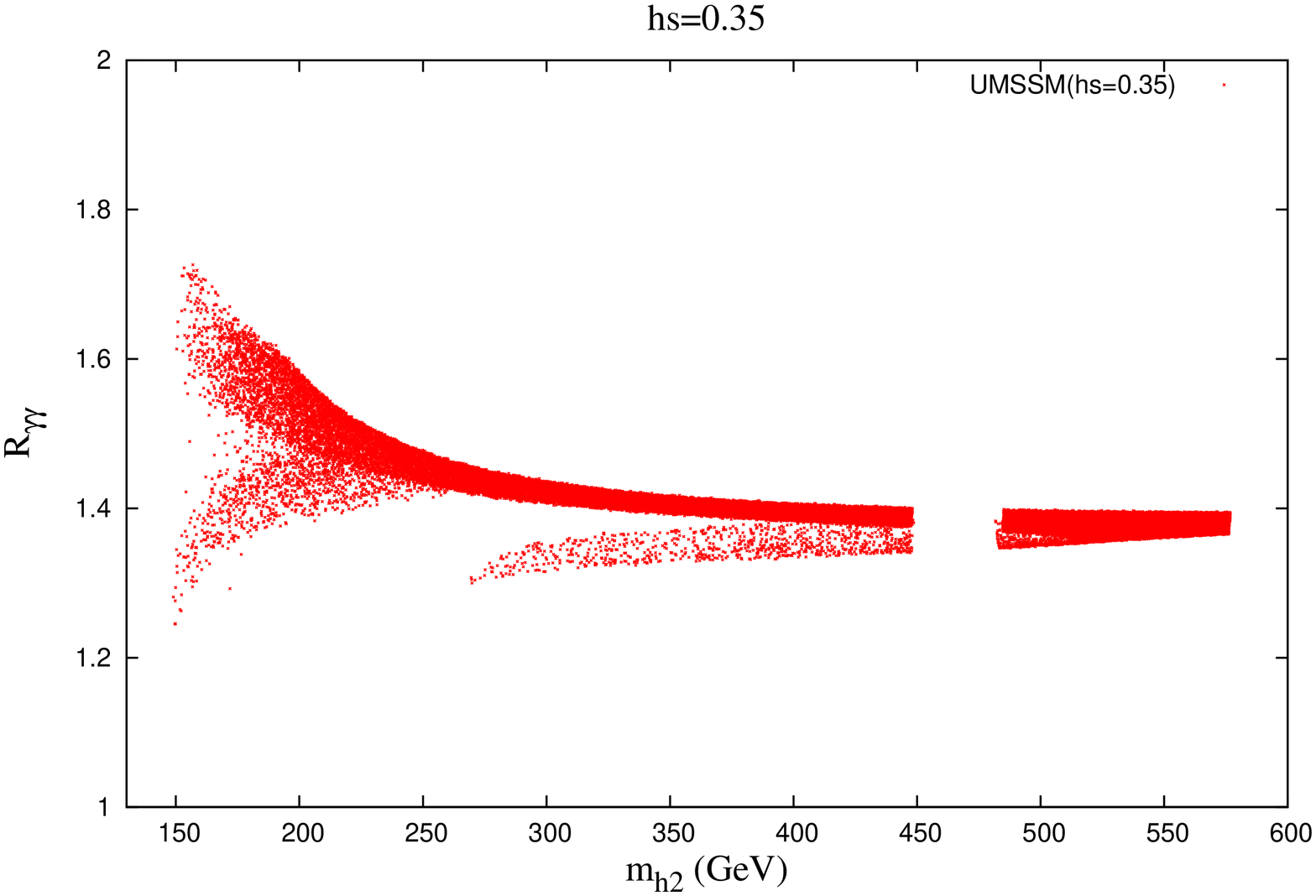}
\includegraphics[angle=270,width=3.2in]{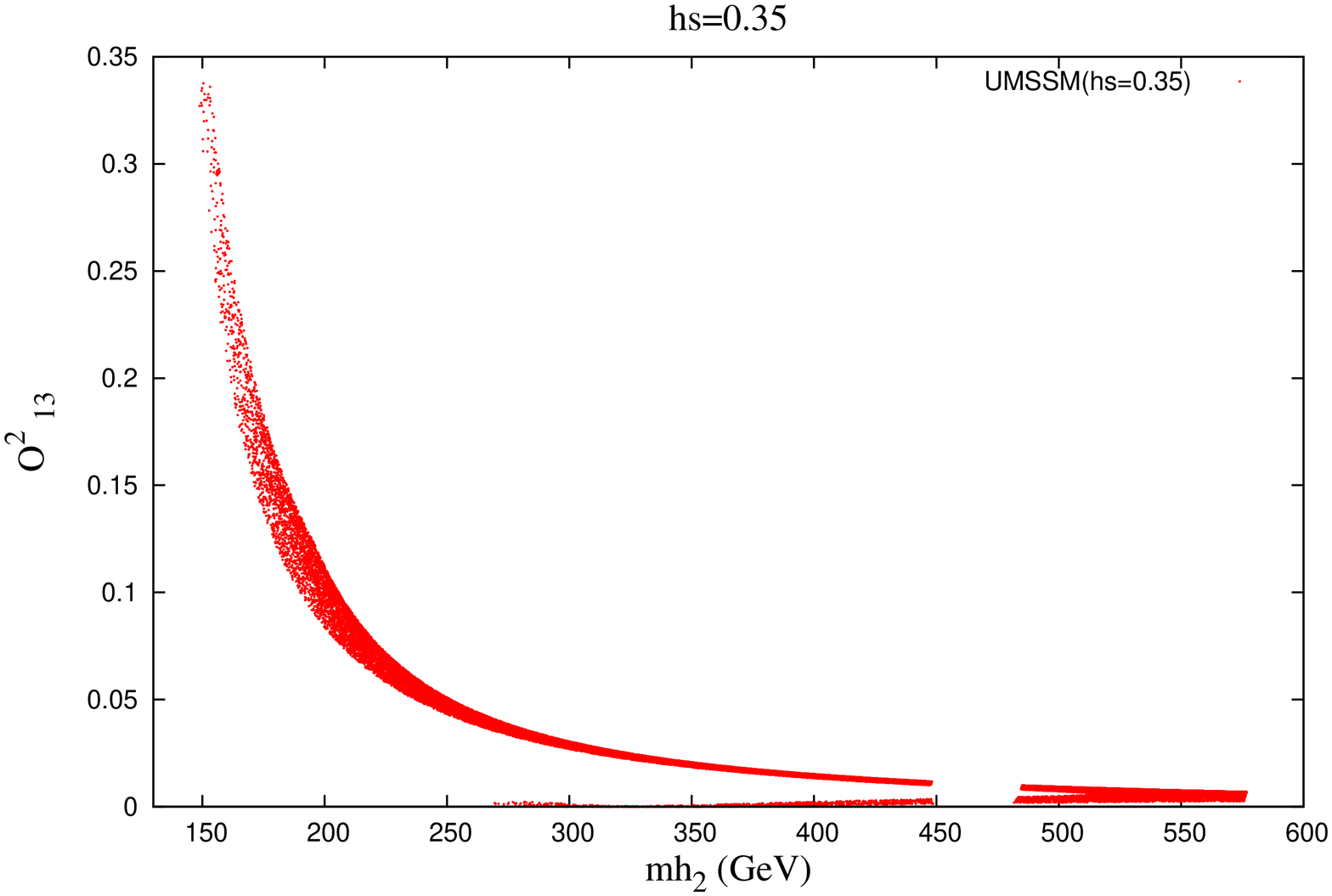}
\caption{\small \label{two}
Same as Fig.~\ref{one}. Case for $h_s = 0.35$. 
}
\end{figure}

\begin{figure}[th!]
\centering
\includegraphics[angle=270,width=3.2in]{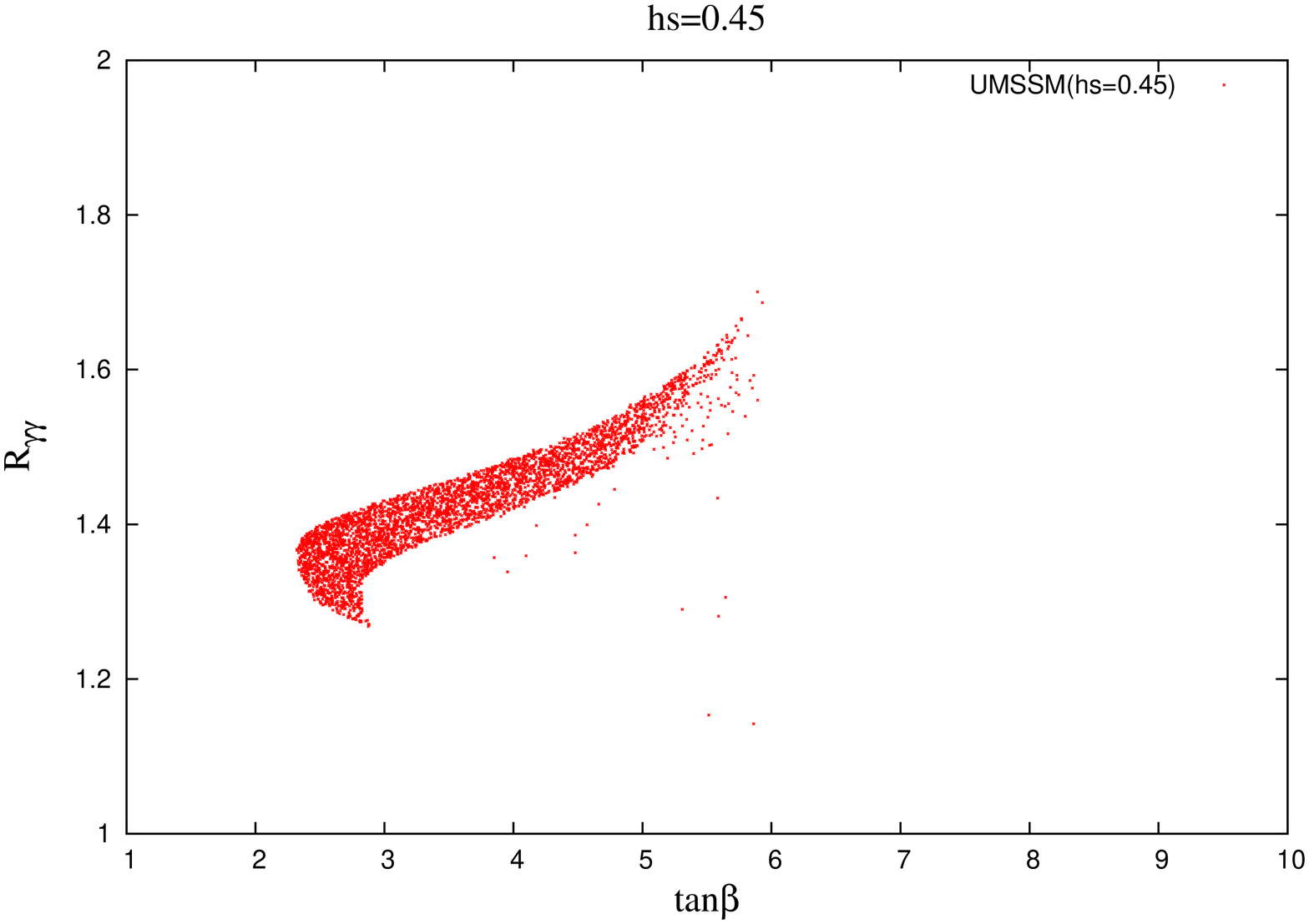}
\includegraphics[angle=270,width=3.2in]{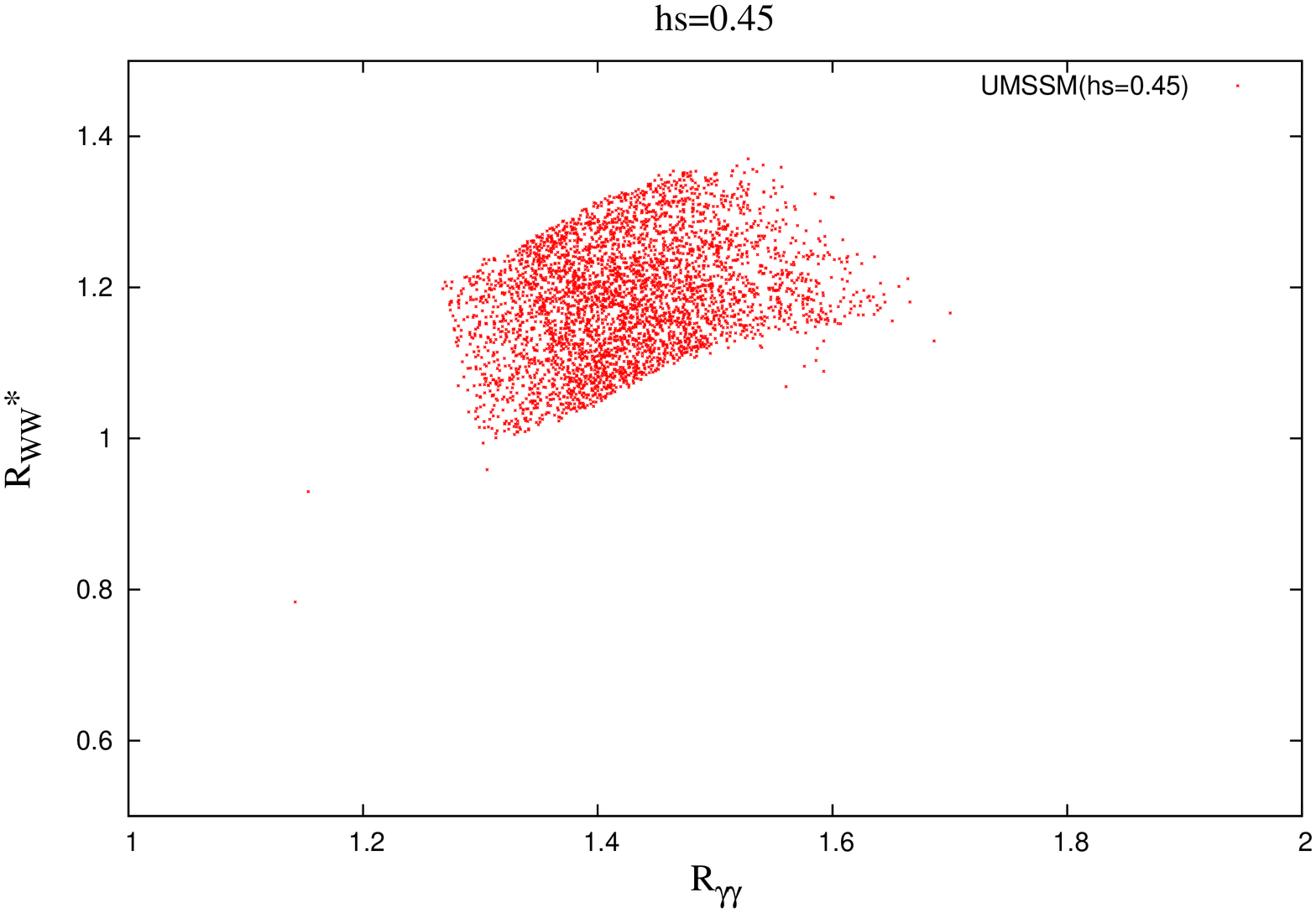}
\includegraphics[angle=270,width=3.2in]{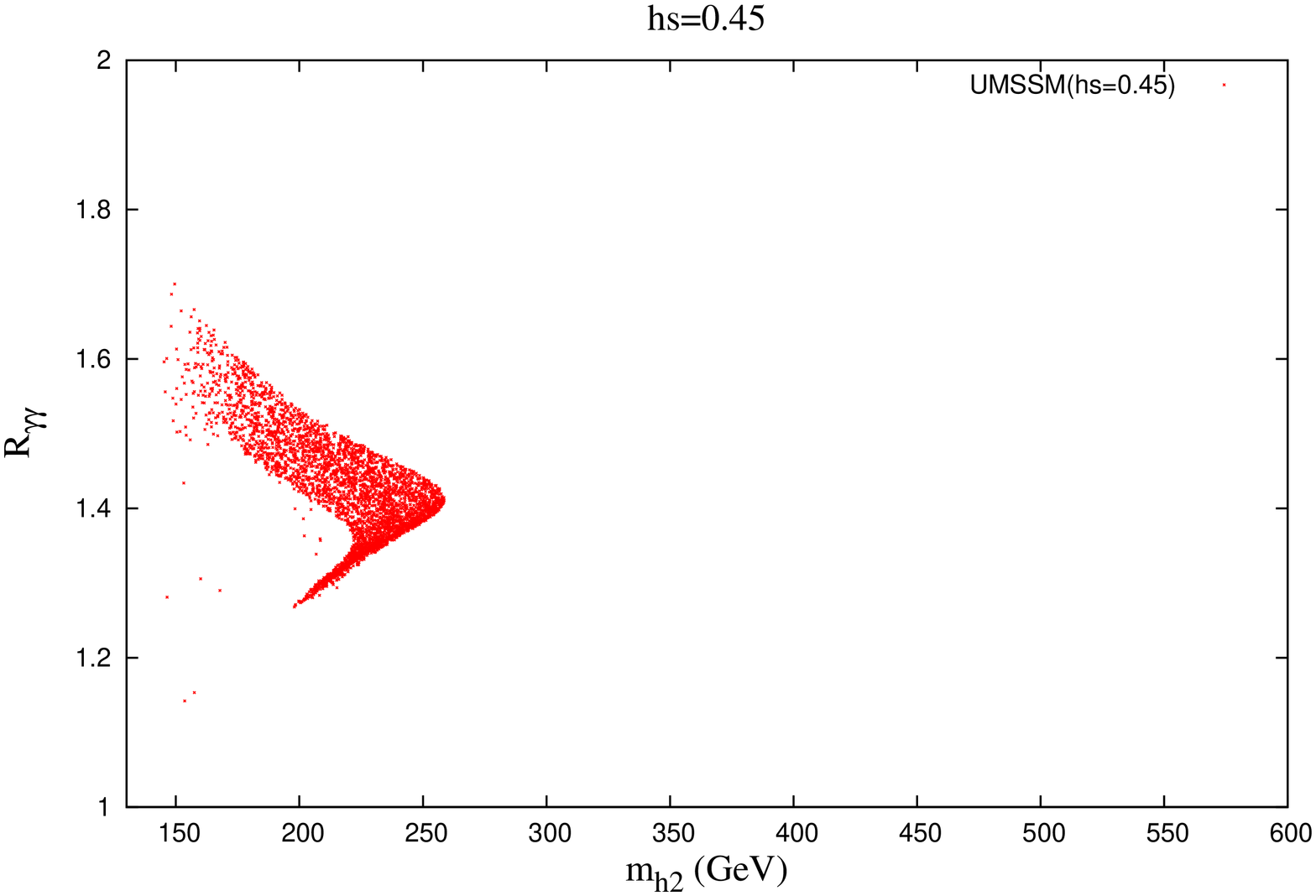}
\includegraphics[angle=270,width=3.2in]{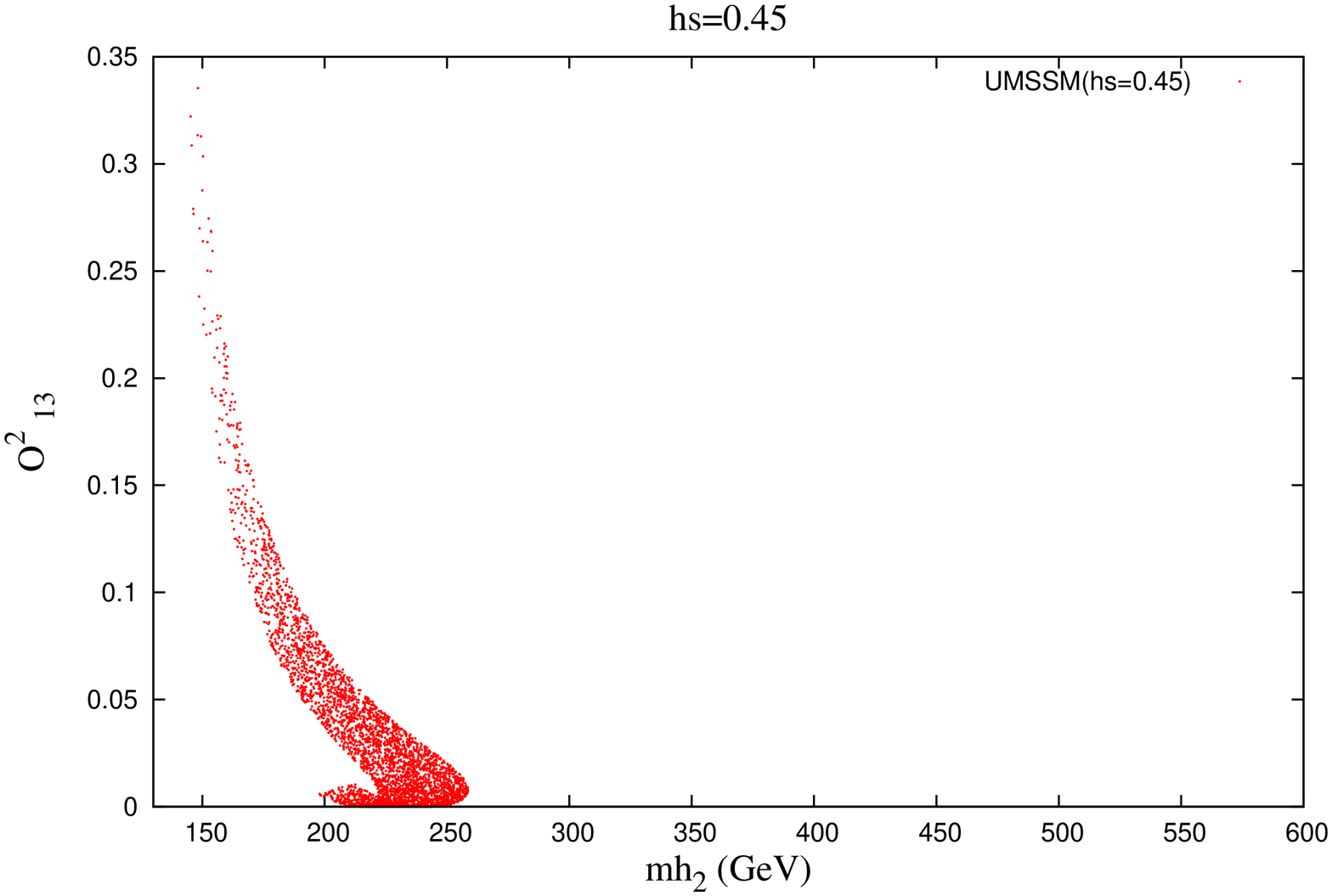}
\caption{\small \label{three}
Same as Fig.~\ref{one}. Case for $h_s = 0.45$. 
}
\end{figure}

\begin{figure}[th!]
\centering
\includegraphics[angle=270,width=3.2in]{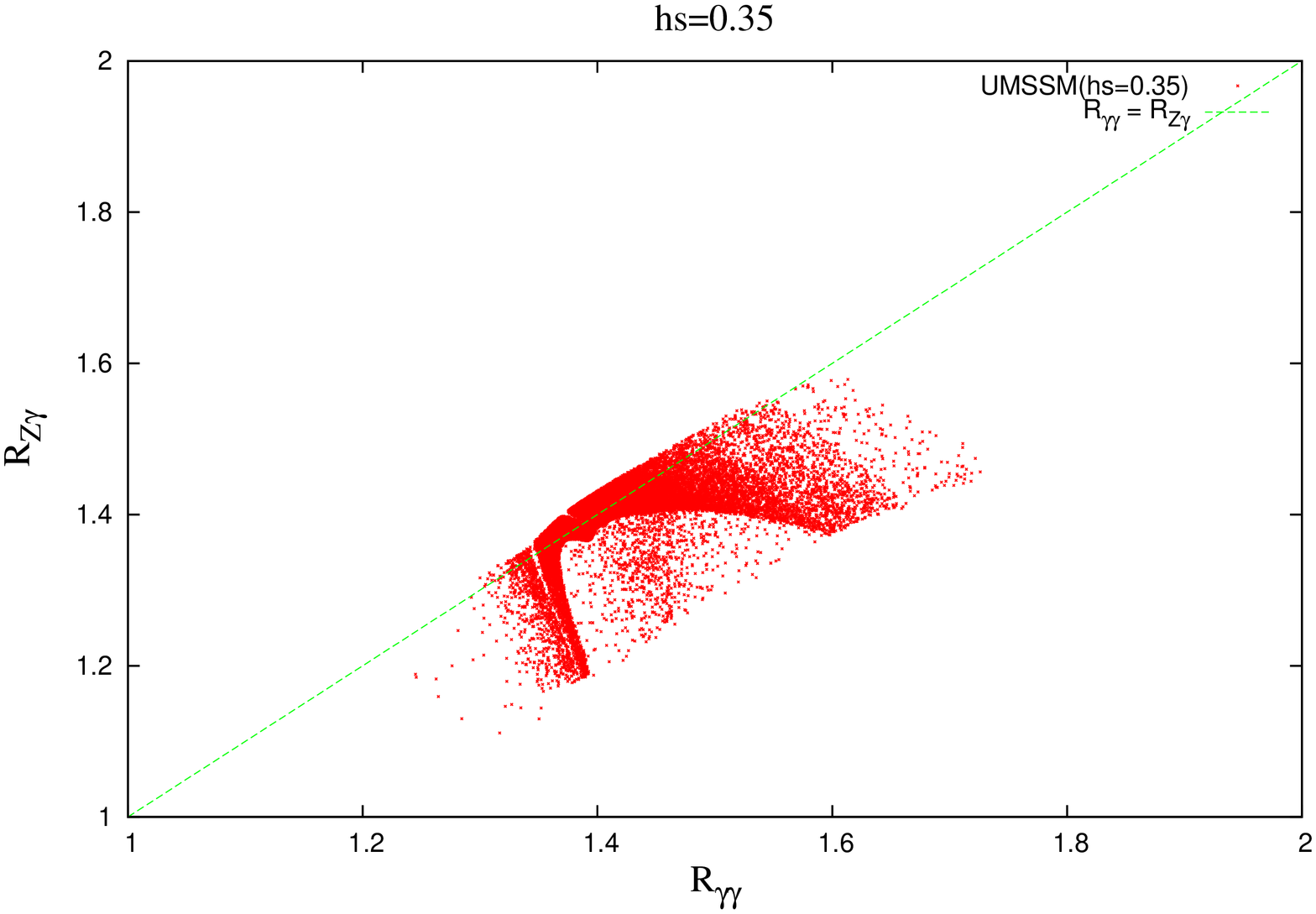}
\includegraphics[angle=270,width=3.2in]{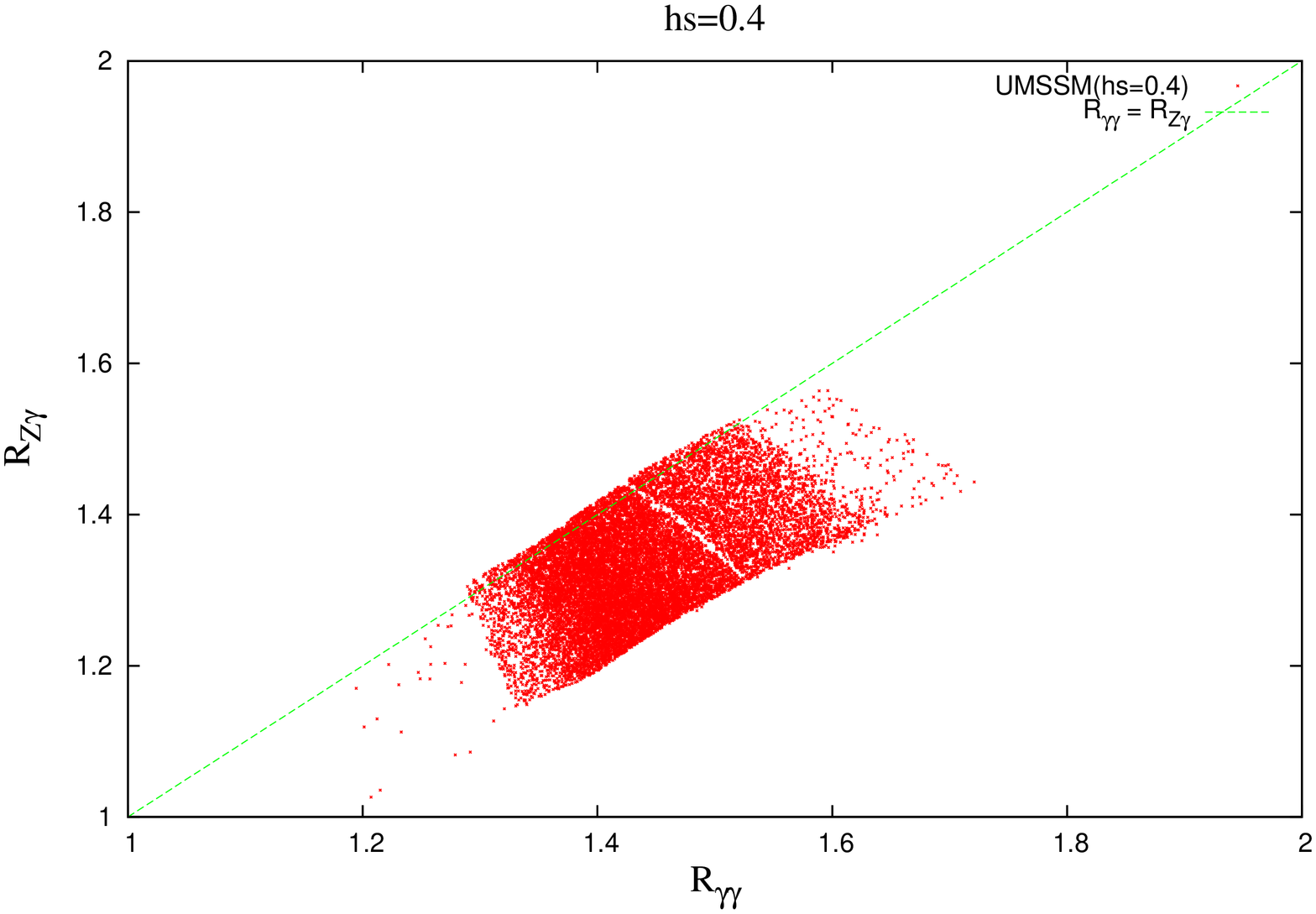}
\includegraphics[angle=270,width=3.2in]{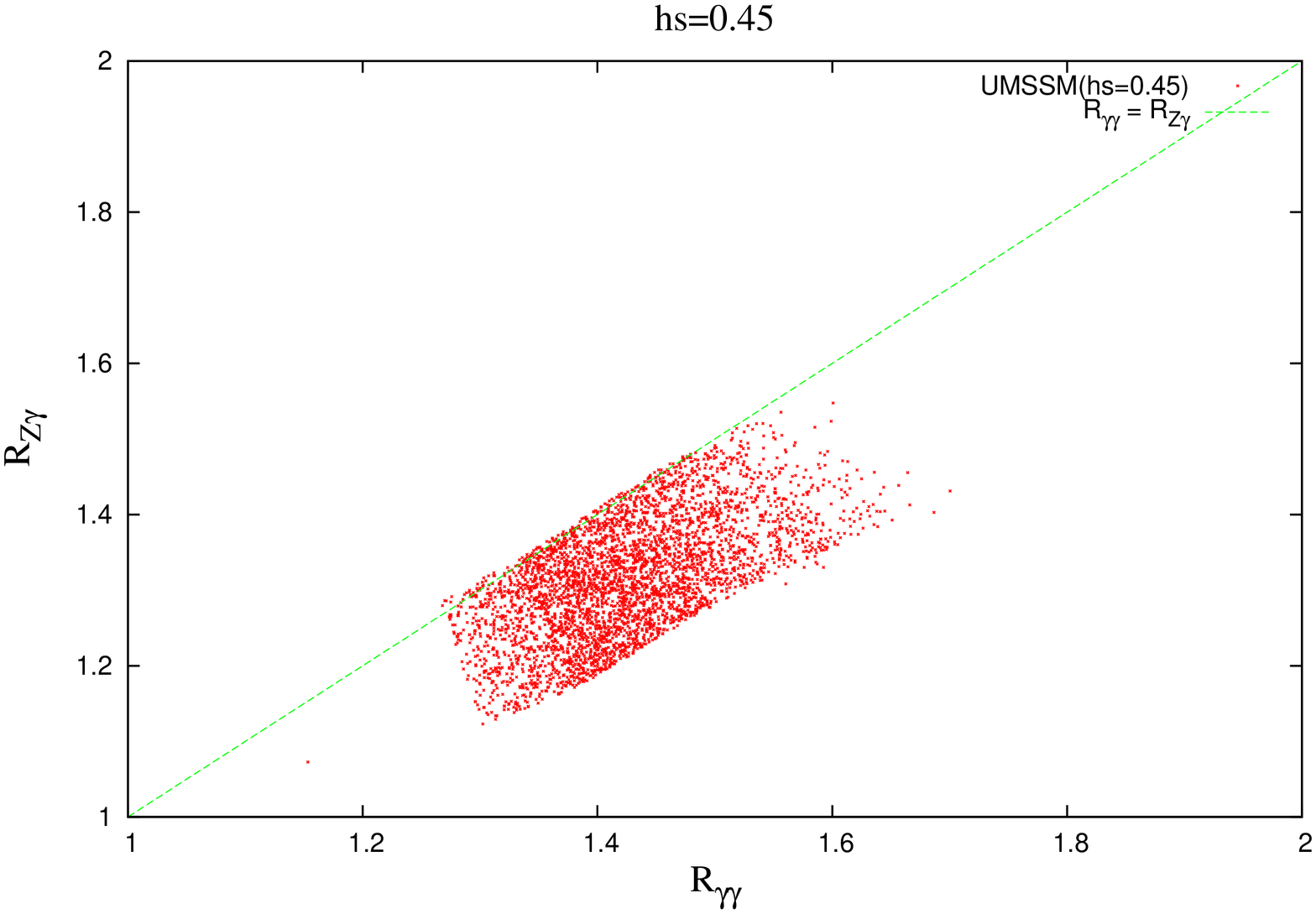}
\caption{\small \label{four}
Correlation between $R_{\gamma\gamma}$ and $R_{Z\gamma}$ for $h_s = 0.35, 0.4,
0.45$.
}
\end{figure}

We start with $h_s = 0.4$ and show relative production rates, as defined by
Eq.~(\ref{R}), in Fig.~\ref{one}. We show $R_{\gamma\gamma}$ versus
$\tan\beta$ in part (a), $R_{WW^*}$ versus $R_{\gamma\gamma}$ in part (b), 
$R_{\gamma\gamma}$ versus $m_{h_2}$ in part (c), and $O_{13}^2$ versus 
$m_{h_2}$ in part (d).
The majority of the points have $R_{\gamma\gamma}$ between $1.3$ and $1.6$ 
while $R_{WW^*}$ (similarly $R_{ZZ^*}$) is 
between $1.0$ and $1.4$ with $\tan\beta$ between 3 and 9.
The correlations between $R_{\gamma\gamma}$ and $m_{h_2}$, and 
between $O_{13}^2$ and $m_{h_2}$ show that the enhancement of $R_{\gamma\gamma}$
of $h_1$ is a result of mixing between the doublet and singlet components.
When $m_{h_2}$ gets closer to $m_{h_1}$, the mixing between $h_1$ and $h_2$ 
gets stronger, and therefore the singlet component $O_{13}^2$ for $h_1$ 
becomes larger and so does $R_{\gamma\gamma}$.
The $R_{\gamma\gamma}$ is enhanced mainly due to a reduced total width, 
which is dominated by the $b\bar b$ width. In order to fully
understand the enhancement of diphotons, we show the branching ratio
$B(h_1 \to b\bar b)$ versus (a) $m_{h_2}$ and (b) $R_{\gamma\gamma}$ in
Fig.~\ref{new2}.  In Fig.~\ref{new2} (a) we can see that the branching
ratio into $b\bar b$ decreases as $m_{h_2}$ approaches $m_{h_1}$,
where the mixing is the strongest. Also, in Fig.~\ref{new2}(b)
$R_{\gamma\gamma}$ increases as $B(h_1 \to b\bar b)$ decreases.  It is
now clear that the enhancement in diphotons is due to a reduced $b\bar
b$ branching ratio, which in turn is because of the stronger mixing
with the singlet.

We repeat the cases of $h_s =0.35$ and $h_s=0.45$ in 
Figs.~\ref{two} and \ref{three}, respectively. 
It is easy to see that the number of points for $h_s=0.35$ 
and $h_s=0.45$ are reduced substantially as compared with $h_s =0.4$.
The range of $\tan\beta$ for $h_s=0.35$ 
stretches between 3 to 40, while for $h_s = 0.45$  
it shrinks drastically to between 2.5 and 6. 
The correlations between $R_{WW^*}$ and $R_{\gamma\gamma}$, 
between $R_{\gamma\gamma}$ and $m_{h_2}$, and between $O_{13}^2$ and $m_{h_2}$ 
are similar to the case of $h_s =0.4$.
Note that there is a gap in $m_{h_2}$ between $450-475$ GeV
in the case of $h_s=0.35$, which is mainly due to the combined
constraints of $R_{\gamma\gamma}$ and $R_{WW^*}$.
We have checked that there are many fewer points satisfying all 
the constraints below $h_s =0.3$ and above $h_s = 0.5$.

An interesting prediction is the relative production rate of $R_{Z\gamma}$,
which can probe various Higgs-sector extensions \cite{cw}.
In the SM, $B(h_{\rm SM} \to Z\gamma)$ is smaller than 
  $B(h_{\rm SM} \to \gamma\gamma)$. 
We show the correlation between $R_{Z\gamma}$ and $R_{\gamma\gamma}$ for 
$h_s=0.35,0.4,0.45$ in Fig.~\ref{four}, in which the points shown 
already satisfy the constraints listed above. All the points that
receive enhancement in the $\gamma\gamma$ channel also receive enhancement in
the $Z\gamma$ channel. However, for most of the points $R_{Z\gamma}$ 
is less than $R_{\gamma\gamma}$, indicated by the points below the green line
($R_{Z\gamma} = R_{\gamma\gamma}$). 

\begin{table}[thb!]
\caption{\small \label{table1}
Selected points (labeled 1, 2, and 3) 
in the allowed parameter space
for $h_s=0.35,0.4,0.45$. The masses are given in GeV. The $h_{\rm SM-like} = h_1$
in our scan.}
\begin{ruledtabular}
\begin{tabular}{l|lll|lll|lll}
& \multicolumn{3}{c|}{$h_s=0.35$} & \multicolumn{3}{c|}{$h_s=0.4$} 
                                 & \multicolumn{3}{c}{$h_s=0.45$} \\
\hline \hline
 & {\sf No. 1} & {\sf No. 2} & {\sf No. 3}   
 & {\sf No. 1} & {\sf No. 2} & {\sf No. 3}  
 & {\sf No. 1} & {\sf No. 2} & {\sf No. 3}  \\
\hline
$m_{h_{\rm SM-like}}$ & 124.42 & 124.02 & 126.04 & 124.02 & 124.01 & 125.45 
   & 124.13 & 125.51 & 124.35 \\ 
$m_{h_{2}}$ & 154.49 & 157.15 & 162.80 & 159.23 & 158.27 & 149.30 
   & 159.59 & 158.88 & 149.48 \\ 
$m_{\tilde{\chi}_{0}}$ & 54.32 & 27.77 & 64.61 & 28.75 & 25.33 & 64.71 
  & 27.84 & 87.70 & 67.11 \\ 
\hline
$ |O_{13}|^{2} $ & 0.316 & 0.296 & 0.210 & 0.248 & 0.256 & 0.319 
   & 0.215 & 0.188 & 0.313 \\ 
$ \tan\beta $ & 19.91 & 21.55 & 19.53 & 9.01 & 9.01 & 8.46 
  & 5.74 & 5.53 & 5.89 \\ 
\hline
 $B(h\rightarrow \gamma\gamma)\times 10^{3} $ & $ 4.23 $ & $ 4.31 $ & $ 3.73 $ 
  & $ 3.83 $ & $ 3.90$ & $ 4.43 $ 
  & $ 3.75 $ & $ 3.62$ & $ 4.41 $ \\ 
$ B(h\rightarrow b\overline{b}) $ & 0.387 & 0.425 & 0.441 
   & 0.444 & 0.443 & 0.400 & 0.480 & 0.473 & 0.428 \\ 
$ B(h\rightarrow \tilde{\chi}^0_1 \tilde{\chi}^0_1 ) $ 
     & 0.054 & 0.007 & 0.0 & 0.039 & 0.033 & 0.0
     & 0.009 & 0.0 & 0.0 \\
\hline
$ R_{\gamma\gamma} $ & 1.63 & 1.72 & 1.61 & 1.63 & 1.64 & 1.66 
  & 1.65 & 1.61 & 1.70 \\ 
$ R_{ZZ^{*}} $ & 1.15 & 1.18 & 1.38 & 1.13 & 1.13 & 1.27 
  & 1.16 & 1.29 & 1.17 \\ 
$ R_{WW^{*}} $ & 1.15 & 1.18 & 1.34 &1.13 & 1.14 & 1.25 
  & 1.16 & 1.26 & 1.17 \\ 
$ R_{Z\gamma} $ & 1.40 & 1.44 & 1.53 & 1.36 & 1.37 & 1.50  
  & 1.39 & 1.47 & 1.43 \\ 
\end{tabular}
\end{ruledtabular}
\end{table}

It is instructional to list a few selected points in the allowed 
parameter space, as shown in Table \ref{table1}. 
The masses $m_{h_{\rm SM-like}}$ are
all around $124-126$ GeV and $m_{h_2}$ are around $150 - 160$ GeV so that
the singlet-doublet mixing is strong but not maximal. The $b\bar b$ width
is reduced by a moderate amount because we have set that the singlet fraction
cannot be too large ($O_{13}^2 < 1/3$).  Therefore, we can
see $R_{WW^*}$ and $R_{ZZ^*}$ are enhanced by about 
$10-15$\%. The $R_{\gamma\gamma}$ is 
enhanced by about 60\% and $R_{Z\gamma}$ by about 40\%. 
In the future, if experiments can measure 
$R_{\gamma\gamma}$, $R_{WW^*}$, and $R_{ZZ^*}$ to better precision,
one could tell whether the enhancement in $R_{\gamma\gamma}$ is due to 
singlet-doublet mixing.

\begin{figure}[th!]
\centering
\includegraphics[angle=270,width=3.2in]{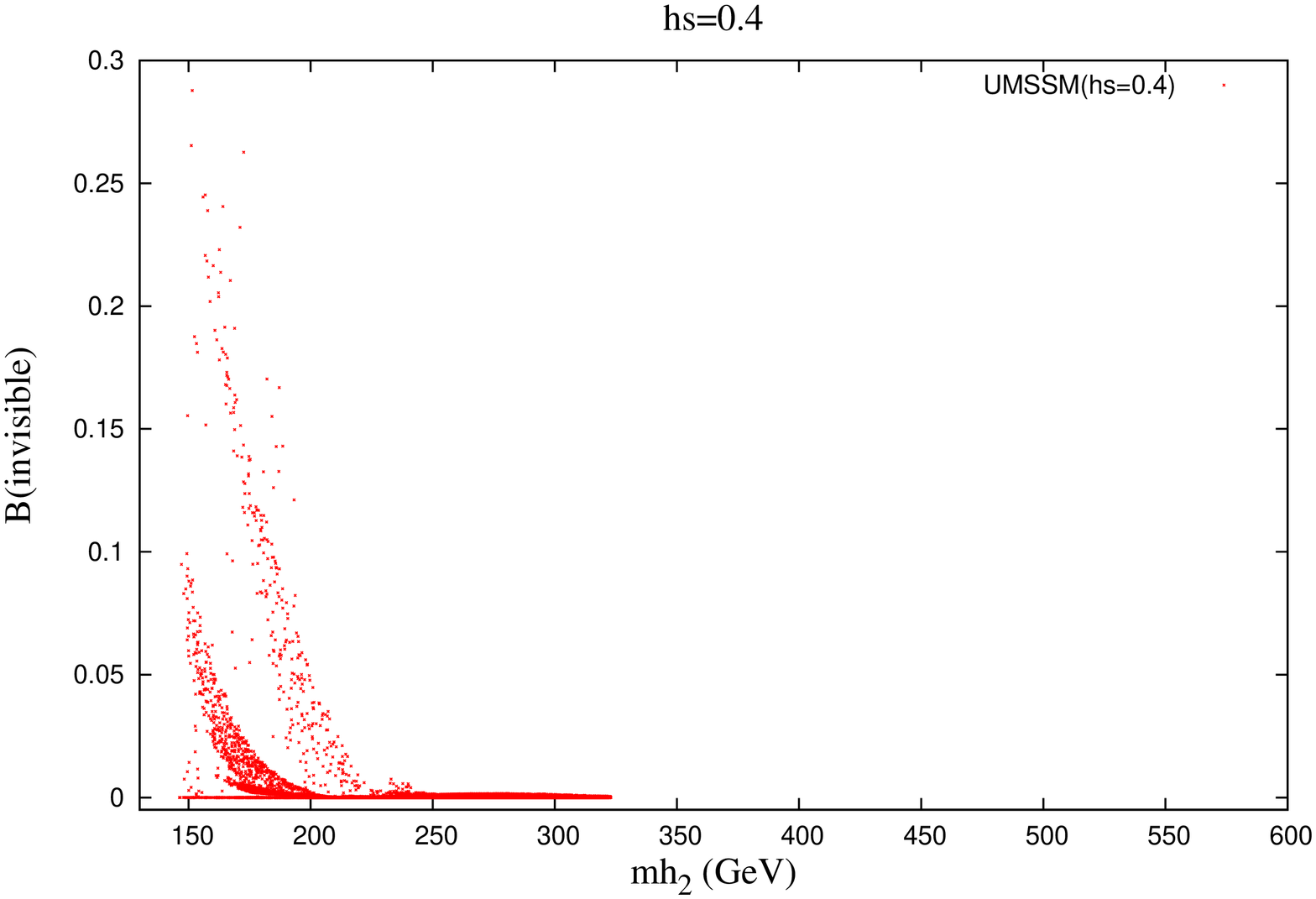}
\includegraphics[angle=270,width=3.2in]{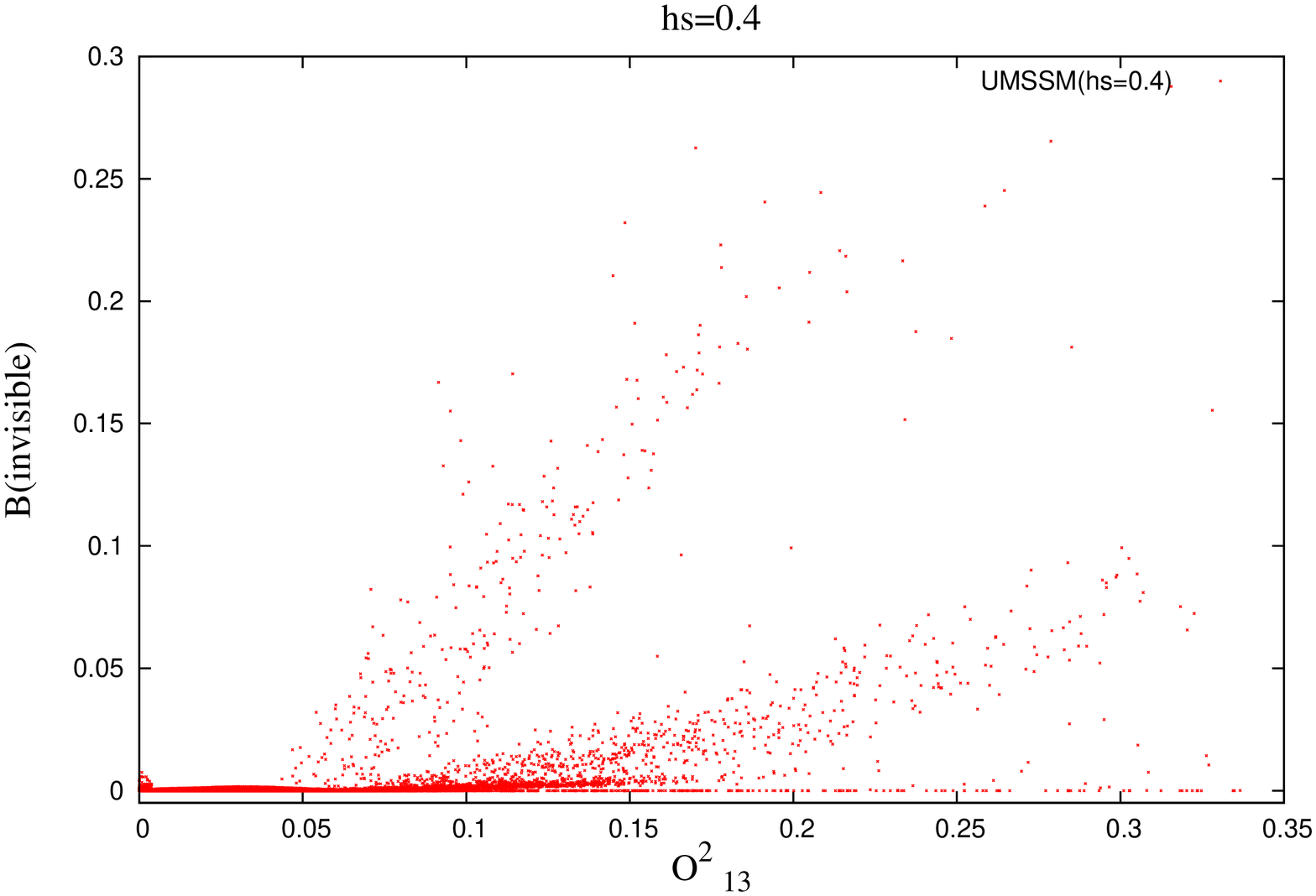}
\caption{\small \label{last}
Case for $h_s=0.4$. Same as Fig.~\ref{one}. Shown is the
invisible branching ratio $B(h_1 \to \tilde{\chi}^0_1 \tilde{\chi}^0_1)$ 
versus (a) $m_{h_2}$ and (b) $O^2_{13}$.
}
\end{figure}

Note that the lightest neutralino $\tilde{\chi}^0_1$ could be lighter
than $m_{h_{\rm SM-like}} /2$. 
In this case $h_1 \to \tilde{\chi}^0_1 \tilde{\chi}^0_1$
is possible, but the branching ratio 
$B(h_1 \to \tilde{\chi}^0_1 \tilde{\chi}^0_1)$ is very small, because we 
have set the production rates $R_{\gamma\gamma}$, $R_{WW^*}$, and $R_{ZZ^*}$ 
larger than certain values. 
In Fig.~\ref{last}, we show the 
branching ratio of the $B(h_1 \to \tilde{\chi}^0_1 \tilde{\chi}^0_1)$ 
(invisible) versus $m_{h_2}$ and $O^2_{13}$. These are the parameter-space
points satisfying all the constraints of chargino mass, invisible $Z$ width,
Higgs boson mass, and the Higgs production rates. We can see that 
the majority of points are at $B({\rm invisible}) =0$, because mostly
$h_1 \to \tilde{\chi}^0_1 \tilde{\chi}^0_1$ is not kinematically open yet;
while the other points have $B({\rm invisible}) \alt  0.25$. 
This is in accord with a recent model-independent study on the Higgs
boson couplings that the nonstandard Higgs decay branching ratio is
constrained to be less than about $0.25$ \cite{global}.

\section{Discussion and Conclusions}

There are two ways to enhance the diphoton production rate, either by
increasing the absolute width into $\gamma\gamma$ or by 
reducing the total width of the Higgs boson (which is dominated by
the 
$b\bar b$ width at 125 GeV). The former is possible if an extra light charged 
particle is running in the triangular loop, e.g., a light stau \cite{carena},
in the MSSM. The latter effect is possible if the SM-like Higgs boson has
a large mixing with another singlet-like Higgs boson, e.g., in the NMSSM
\cite{ellw}, such that the $b\bar b$ width is reduced by the mixing and 
therefore the $\gamma\gamma$ branching ratio is enhanced. 

For the choice of UMSSM parameters all the extra charged particles like 
the stau, top squark, sbottom and the charged Higgs boson are relatively heavy.
We have searched in the parameter space of UMSSM under the constraints of
current Higgs boson data, chargino-mass bound, and $Z$ invisible width.
We found that 
(1) the enhancement of the diphoton production rate is mainly due to the mixing
between the Higgs doublets and singlet, and 
(2) the lightest CP-even Higgs boson is SM-like while the second
lightest is more singletlike. This is in contrast to the case of NMSSM,
in which the lightest is singletlike and the second lightest is SM-like.

Before closing, we offer a few more comments as follows.

\begin{enumerate}
\item
The relative production rate $R_{Z\gamma}$ mostly goes in the same 
direction as $R_{\gamma\gamma}$, though the amount of enhancement in
$R_{Z\gamma}$ is less than $R_{\gamma\gamma}$. The probing of the $Z\gamma$
mode of the observed Higgs boson is an interesting test for the Higgs
boson from the SM or from its extensions. In the present luminosity, it is
rather difficult to probe the $Z\gamma$ because it suffers an 
additional suppression from the leptonic branching ratio of the $Z$
boson.

\item 
Almost all of the points have $R_{WW^*}$ between $1.0$ and $1.4$. This is
easy to understand because $R_{\gamma\gamma}$ is enhanced by a reduced total
width.  Therefore, the $WW^*$ and $ZZ^*$ branching ratios also increase.

\item
The mass of the second lightest CP-even Higgs boson cannot be too large,
as shown in bottom panels 
of Figs.~\ref{one}, \ref{two}, \ref{three}. Again, this is easy
to understand because in order to achieve a large doublet-singlet mixing
between $h_1$ and $h_2$ their mass difference cannot be too large. We found
that $m_{h_2} < 580, 320, 260$ GeV for $h_s =0.35, 0.4, 0.45$ respectively.
The detection of $h_2$ is rather difficult because of its singlet nature.
The production cross section would be reduced significantly by the mixing.

\item There are six physical neutralinos in the mass spectrum in
  UMSSM. The lightest one can be the dark matter candidate. If
  kinematics is allowed, it may lead to invisible modes for the decays
  of Higgs bosons, $Z'$ or even $Z$. Dark matter physics is therefore
  very rich in this model. We only touch upon this lightly in this
  work and would like to return to this issue in future publications.

\end{enumerate}

\newpage

\appendix
\section{Loop Functions}

The partial decay width for $h_j \to \gamma\gamma$ is given by Eq.(\ref{hgam}). 
The loop functions are given by
\begin{eqnarray}
\label{Ff}
F_{f}&=& -2 x_{f}\left[1+\left(1-x_{f}\right)f(x_{f})\right]  R_{f}  \qquad \qquad (f = \tau, t, b)\\
F_{W} &=& \left[2+3x_{W} +3x_{W}\left(2-x_{W}\right)f(x_{W}) \right] R_{W} \\
F_{h^{\pm}} &=& x_{h^{\pm}} \left[ 1-x_{h^{\pm}}f(x_{h^{\pm}})\right]
   R_{h^{\pm}}  \frac{m_{W}^{2}}{m_{h^{\pm}}^{2}}  
\end{eqnarray}
for non-SUSY particles and
\begin{eqnarray}
\label{Fsf}
F_{\tilde f}&=& \sum_{i=1,2} x_{\tilde{f _{i}}}\left[1-x_{\tilde{f _{i}}}f(x_{\tilde{f _{i}}})\right]
R_{h_j \tilde{f _{i}}\tilde{f _{i}}} \frac{m_{Z}^{2}}{m_{\tilde{f _{i}}}^{2}} \qquad \qquad (\tilde f = \tilde \tau, \tilde t,\tilde b) \\
F_{\tilde \chi^{\pm}}&=& \sum_{i=1,2} -2 x_{\tilde{\chi}^{\pm}_{i}}
 \left[ 1+ \left(1-x_{\tilde{\chi}^{\pm}_{i}}\right) f(x_{\tilde{\chi}^{\pm}_{i}})
 \right ]  R_{\tilde{\chi}^{\pm}_{i}}  \frac{m_{W}}{m_{\tilde \chi^{\pm}_i} } 
\end{eqnarray}
for sparticles with $x_{X}  =4 m_{X}^{2} / m_{h_j}^{2} \, (X = \tau, t, b, W, h^\pm , \tilde \tau_i, \tilde t_i, \tilde b_i , \tilde \chi_i^\pm)$.
\begin{equation}
\label{f}
f \left( x \right) =
\begin{cases} \left[ \sin^{-1} \sqrt \frac{1}{x} \right]^2  & \;\; \mbox{for } x \ge 1 \\ 
-\frac{1}{4} \left[ \ln  \left(\frac{1 + \sqrt{1 - x}}{1 - \sqrt{1 - x}} \right)
- i \pi\right]^2 & \;\; \mbox{for } x < 1 
\end{cases} 
\end{equation}
The couplings entering into the loop functions for the non-SUSY particles are
\begin{equation}
 R_\tau = \frac{O_{j1}}{\cos\beta}  \;\; , \;\; R_{t} = \frac{O_{j2}}{\sin\beta}  \;\; , \;\; R_b = R_\tau
\end{equation} 
\begin{eqnarray} 
 R_{W} &=& O_{j2} \sin\beta + O_{j1} \cos\beta \\
 R_{h^{\pm}} &=& \frac{3-2\sin^{2}\theta_{W}}{2\cos^{2}\theta_{W}} \sin\beta \cos\beta 
  \left( O_{j2} \cos\beta +O_{j1} \sin\beta \right) \nonumber \\
&&  
  +\frac{1-2\sin^{2}\theta_{W}}{2\cos^{2}\theta_{W}}\left( O_{j2} \sin^{3}\beta +O_{j1} \cos^{3}\beta \right)
 \nonumber \\
&&
  +\frac{2g_{2}^{2}}{g^{2}}Q'_{H_u} {}^2 O_{j2} \sin\beta
 \cos^{2}\beta +\frac{2g_{2}^{2}}{g^{2}}Q'_{H_d} {}^{2} O_{j1} \sin^{2}\beta
 \cos\beta  \nonumber \\
&&
    +\frac{2g_{2}^{2}}{g^{2}}Q'_{H_u}Q'_{H_d} 
  \left( O_{j2} \sin^{3}\beta +O_{j1}\cos^{3}\beta \right)  \nonumber \\
&&
  -\frac{2h_{s}^{2}}{g^{2}} \sin\beta \cos\beta 
 \left( O_{j2} \cos\beta + O_{j1} \sin\beta  \right) \nonumber \\
&&  
+ \left(\frac{h_{s}^{2} v_{s}}{gm_{W}}
+ \frac{g_{2}^{2}}{gm_{W}}Q'_{H_u}Q'_{S} v_s \cos^2\beta    
  +\frac{g_{2}^2}{gm_{W}}Q'_{H_d}Q'_{S} v_s \sin^2\beta  \right. \nonumber \\
&& \left. \qquad
  + \; \frac{\sqrt{2}h_{s}A_{s}} {gm_{W}} \sin\beta \cos\beta \right) O_{j3} 
\end{eqnarray}
For the sfermions, we have the couplings
\begin{eqnarray}
R_{h_j \tilde{f_{1}}\tilde{f_{1}}} &=& R^L_{\tilde f} \cos^{2}\theta_{\tilde f}
+ R^R_{\tilde f} \sin^{2} \theta_{\tilde f}+2 R^{RL}_{\tilde f} \sin\theta_{\tilde f} \cos\theta_{\tilde f} 
\\
R_{h_j \tilde{f_{2}}\tilde{f_{2}}} &=& R^L_{\tilde f} \sin^{2}\theta_{\tilde f}
+ R^R_{\tilde f} \cos^{2}\theta_{\tilde f} - 2 R^{RL}_{\tilde f} \sin\theta_{\tilde f} \cos\theta_{\tilde f} 
\end{eqnarray}
where $\theta_{\tilde f}$ is the mixing angle 
between $\tilde f_L$ and $\tilde f_R$ to obtain 
the physical mass eigenstates $\tilde f_1$ and $\tilde f_2$. 
For the $Z\gamma$ case, we also need the off-diagonal term
\begin{equation}
R_{h_j \tilde{f_{1}}\tilde{f_{2}}} = \left( R^R_{\tilde f} - R^L_{\tilde f}\right) \sin\theta_{\tilde f}\cos\theta_{\tilde f}
+R^{RL}_{\tilde f} \left(\cos^{2}\theta_{\tilde f}-\sin^{2}\theta_{\tilde f}\right) \; .
\end{equation}
The expressions of $R^{L,R,RL}_{\tilde t , \tilde b}$ are given by
\begin{eqnarray}    
 R^L_{\tilde t , \tilde b} &=& \frac{v m_{W}}{gm_{Z}^{2}} \Biggr [
 \left(\frac{g^{2}}{2\cos^{2}\theta_{W}} \left( \sin^{2}\theta_{W} Q^{t,b} - T^{t,b}_3 \right) +
  g_{2}^{2}Q'_{H_u}Q'_{Q_3} \right)  O_{j2} \sin\beta +
 \frac{2m_{t,b}^{2}}{v^2} R_{t,b} \nonumber\\
&& + 
 \left(-\frac{g^{2}}{2\cos^{2}\theta_{W}} \left( \sin^{2}\theta_{W} Q^{t,b} - T^{t,b}_3 \right) +
 g_{2}^{2}Q'_{H_d}Q'_{Q_3}\right)  O_{j1} \cos\beta  \nonumber \\
 && + \; g_{2}^{2} \frac{v_s}{v} Q'_{S}Q'_{Q_3}  O_{j3}   \Biggr] \\
 R^R_{\tilde t , \tilde b} &=&\frac{v m_{W}}{gm_{Z}^{2}}\Biggr[
 \left(-\frac{g^{2}\sin^{2}\theta_{W}}{2\cos^{2}\theta_{W}}Q^{t,b}+g_{2}^{2}Q'_{H_u}Q'_{U_3^{\rm c},D_3^{\rm c}} \right)
   O_{j2} \sin\beta +\frac{2m_{t,b}^{2}}{ v^2} R_{t,b} \nonumber\\
 && +
 \left(\frac{g^{2}\sin^{2}\theta_{W}}{2\cos^{2}\theta_{W}}Q^{t,b}+g_{2}^{2}Q'_{H_d}Q'_{U_3^{\rm c},D_3^{\rm c}} \right) 
    O_{j1} \cos\beta +g_{2}^{2} \frac{v_s}{v} Q'_{S}Q'_{U_3^{\rm c},D_3^{\rm c}}  O_{j3} \Biggr ] \\
R^{RL}_{\tilde t , \tilde b} &=&\frac{v m_{t,b}}{2m^{2}_{Z}}\left(\frac{g}{2m_{W}}A_{t,b}R_{t,b}
-h_{s}\left[\frac{gv_{s}}{2\sqrt{2}m_{W}}R'_{t,b}-\frac{1}{\sqrt{2}} R^{''}_{t,b} \right]\right) 
\end{eqnarray}
where we have defined 
\begin{equation}
 R'_\tau = \frac{O_{j2}}{\cos\beta}  \;\; , \;\; R'_{t} = \frac{O_{j1}}{\sin\beta}  \;\; , \;\; R'_b = R'_\tau
\end{equation} 
\begin{equation}
 R^{''}_\tau = O_{j3} \tan\beta  \;\; , \;\; R^{''}_{t} = O_{j3} \cot\beta  \;\; , \;\; R^{''}_b = R^{''}_\tau
\end{equation} 
The $R^{L,R,RL}_{\tilde \tau}$ can be obtained from the $R^{L,R,RL}_{\tilde b}$ by appropriate
substitutions.

For the chargino loop, we have
\begin{equation}
 R_{\tilde{\chi}^{\pm}_{i}} = 2 \left[\frac{1}{\sqrt{2}}V_{i1}U_{i2}O_{j1}
   +\frac{1}{\sqrt{2}}V_{i2}U_{i1}O_{j2}
  +\frac{h_{s}}{\sqrt{2}g}V_{i2}U_{i2}O_{j3} \right] \; ,
\end{equation}
where $U$ and $V$ are the two unitary matrices that diagonalize 
the chargino mass matrix.

The partial decay width for $h_j \to Z \gamma$ is given by Eq.(\ref{hZgam}).
The loop functions for the non-SUSY particles are given by 
\begin{equation}
G_{f} = N^f_C \cdot R_{f} \cdot \frac{-2 Q^f \left[ T^f_3 -2 Q^f \sin^{2}\theta_{W}\right]}{\sin\theta_{W}\cos\theta_{W}}
\left[I_{1}(\tau _{f},\lambda _{f})-I_{2}(\tau _{f},\lambda _{f})\right]  \qquad (f = \tau , t , b)
\end{equation}
\begin{eqnarray}
G_{W} &=&-R_{W} \cot\theta_{W}\biggl(4\left(3-\tan^{2}\theta_{W}\right)I_{2}(\tau _{W},\lambda _{W}) \biggr. \nonumber \\
&& \left. +\left[ \left(1+\frac{2}{\tau _{W}}\right)\tan^{2}\theta_{W}
-\left(5+\frac{2}{\tau _{W}}\right)\right]I_{1}(\tau _{W},\lambda _{W}) \right)
 \nonumber \\ \\
G_{h^{\pm}} &=&R_{h^{\pm}}\frac{1-2\sin^{2}\theta_{W}}{\sin\theta_{W}\cos\theta_{W}}I_{1}(\tau _{h^{\pm}},\lambda _{h^{\pm}})\frac{m_{W}^{2}}{m_{h^{\pm}}^{2}} 
\end{eqnarray}
Here, we define
\begin{equation}
\tau_X = \frac{4m_X^2}{m_{h_j}^2} \; \; , \; \; \lambda_X = \frac{4 m_X^2}{m_Z^2} 
\qquad \qquad (X = \tau, t, b, W, h^\pm) \; .
\end{equation}
The definitions of $ I_{1}(\tau ,\lambda)$ and $I_{2}(\tau,\lambda)$
are the same as given in \cite{hunter}.
\begin{eqnarray}
I_1(\tau, \lambda) & = & 
\frac{\tau \lambda}{2\left( \tau - \lambda\right)} 
+ \frac{\tau^2 \lambda^2}{2\left( \tau - \lambda\right)^2}\left[ f(\tau) - f(\lambda) \right]
+ \frac{\tau^2 \lambda}{\left( \tau - \lambda\right)^2}\left[ g(\tau) - g(\lambda) \right]
\\
I_2(\tau, \lambda) & = & - \frac{\tau \lambda}{2\left( \tau - \lambda\right)} \left[ f(\tau) - f(\lambda) \right]
\end{eqnarray}
where $f(x)$ is given in Eq.(\ref{f}) and $g(x)$ is defined as
\begin{equation}
\label{g}
g \left( x \right) =
\begin{cases} \sqrt{x - 1} \left[ \sin^{-1} \sqrt \frac{1}{x} \right]  & \;\; \mbox{for } x \ge 1 \\ 
\frac{1}{2} \sqrt{1 - x} \left[ \ln  \left(\frac{1 + \sqrt{1 - x}}{1 - \sqrt{1 - x}} \right)
- i \pi\right] & \;\; \mbox{for } x < 1 
\end{cases} 
\end{equation}
For the sparticles, we have
\begin{eqnarray}
G_{\tilde f} &=& 8 \cdot N^f_C \cdot Q^f \cdot m_{Z}^{2}
\sum_{k,l=1,2}R_{h_j \tilde{f_{l}}\tilde{f_{k}}}R_{Z\tilde{f_{k}}\tilde{f_{l}}}
C_{2}(m_{\tilde{f_{l}}},m_{\tilde{f_{k}}},m_{\tilde{f_{k}}}) 
\\
G_{\tilde \chi ^{\pm}} &=&\sum_{k,l=1,2}\frac{m_{Z}m_{\tilde \chi ^{+}_{l}}}{\sin\theta_{W}}
f\left(m_{\tilde \chi ^{+}_{l}},m_{\tilde \chi ^{+}_{k}},m_{\tilde \chi ^{+}_{k}}\right)
\sum_{m,n=L,R}R^{m}_{Z\tilde \chi ^{+}_{l} \tilde \chi ^{-}_{k}}R^{n}_{h_j \tilde \chi ^{+}_{k} \tilde \chi ^{-}_{l}} 
\end{eqnarray}
The definitions of $C_{2}(m_{1},m_{2},m_{2})$ and $ f(m_{1},m_{2},m_{2})$
can be found in \cite{Djouadi:1996yq}. The couplings for the sfermions are
\begin{eqnarray}
R_{Z\tilde{f_{1}}\tilde{f_{1}}} &=&\frac{1}{\sin\theta_{W}\cos\theta_{W}}
\left[ \left( T^f_3 - Q^f \sin^{2}\theta_{W} \right)\cos^{2}\theta_{\tilde f}
-Q^f\sin^{2}\theta_{W}\sin^{2}\theta_{\tilde f}\right] 
\\
R_{Z\tilde{f_{2}}\tilde{f_{2}}} &=&\frac{1}{\sin\theta_{W}\cos\theta_{W}}
\left[ - Q^f \sin^{2}\theta_{W} \cos^{2}\theta_{\tilde f}
+\! \left( T^f_3 - Q^f  \sin^{2}\theta_{W} \right) 
\sin^{2}\theta_{\tilde f} \right] 
\\
R_{Z\tilde{f_{1}}\tilde{f_{2}}} &=&\frac{-T^f_3}{\sin\theta_{W}\cos\theta_{W}}\sin\theta_{\tilde f}
\cos\theta_{\tilde f} 
\end{eqnarray}
For the charginos, the couplings are
\begin{eqnarray}
R^{L}_{Z\tilde\chi ^{+}_{l}\tilde\chi ^{-}_{k}} &=&-\left(V_{l1}V_{k1}+\frac{1}{2}V_{l2}V_{k2}-\delta_{lk}\sin^{2}\theta_{W}\right) 
\\
R^{R}_{Z\tilde \chi ^{+}_{l}\tilde\chi ^{-}_{k}} &=&-\left(U_{l1}U_{k1}+\frac{1}{2}U_{l2}U_{k2}-\delta_{lk}\sin^{2}\theta_{W}\right)
\\
R^{L}_{h_j \tilde\chi ^{+}_{i}\tilde\chi ^{-}_{l}} &=&\frac{1}{\sqrt{2}}\left[V_{l1}U_{i2}O_{j 1}+V_{l2}U_{i1}O_{j 2}
+\frac{h_{s}}{g}V_{l2}U_{i2}O_{j 3}\right]
\\
R^{R}_{h_j \tilde\chi ^{+}_{i}\tilde\chi ^{-}_{l}} &=&\frac{1}{\sqrt{2}}\left[V_{i1}U_{l2}O_{j 1}+V_{i2}U_{l1}O_{j 2}
+\frac{h_{s}}{g}V_{i2}U_{l2}O_{j 3}\right]
\end{eqnarray}

The partial decay width for $h_j \to gg$ 
is given by Eq.(\ref{hglue}). The loop functions $F_f$ and $F_{\tilde f}$ 
for the colored particles are the same as in the case of 
$h_j \to \gamma \gamma$ given by
Eqs.(\ref{Ff}) and (\ref{Fsf}) respectively.


\section*{Acknowledgments}
This work was supported in parts by the National Science Council of
Taiwan under Grants No. 99-2112-M-007-005-MY3 and
No. 101-2112-M-001-005-MY3, 
and the WCU program through the KOSEF
funded by the MEST (R31-2008-000-10057-0).
TCY is grateful to the National Center for Theoretical Sciences of Taiwan 
for its warm hospitality. 

%

\end{document}